\documentclass[12pt]{article}

\usepackage{newtxtext,newtxmath}

\DeclareUnicodeCharacter{2009}{\,}
\usepackage{graphicx}

\usepackage[letterpaper,margin=1in]{geometry}

\usepackage{booktabs}

\renewenvironment{abstract}
	{\quotation}
	{\endquotation}

\date{}

\makeatletter
\renewcommand{\fnum@figure}{\textbf{Figure \thefigure}}
\renewcommand{\fnum@table}{\textbf{Table \thetable}}
\makeatother

\usepackage{scicite}

\usepackage{siunitx}

\usepackage{url}

\def\scititle{
	The interplay between topology, defects and chiral order in the nearly-commensurate charge density wave of 1T-TaS$_2$
}
\title{\bfseries \boldmath \scititle}

\author{
	Michael Verhage$^{1\ast}$,
	Martin Lee$^{1}$,
	Kees Flipse$^{1\ast}$\and
	$^{1}${Eindhoven University of Technology, Applied Physics and Science
    Education}\\
	\small$^\ast$Corresponding authors. Email: m.verhage@tue.nl, c.f.j.flipse@tue.nl
}

\begin{document} 


\maketitle

\begin{abstract}
\textbf{In 1\textit{T}-TaS$_2$, the nearly-commensurate charge density wave (NC-CDW) exhibits  emergent chirality, a property of interest for technological applications such as memory. We study the relationship between chirality and topological defects using holographic analysis of scanning tunneling microscopy data. By characterizing the CDW order parameter, we uncover distinct topological defects and define a chiral order parameter that connects directly to them. Our analysis distinguishes between vortex pairs in high-strain areas and glass-like soliton or discommensurations networks. We propose that perturbations, such as strain, drive vortex nucleation and annihilation, leaving solitons or discommensurations pinned by disorder. Electric fields induce soliton and discommensurations movement, causing chiral switching via localized order parameter melting, enabling controllable room-temperature chirality.}
\end{abstract}

\newpage
\noindent
\section*{Introduction}

Chirality, or `handedness', represents a fundamental symmetry-breaking phenomenon where an object cannot be superimposed onto its mirror image. This geometric property has potentially important implications in quantum materials, governing phenomena such as the giant anomalous Hall effect \cite{Cao2025Nonlinear2}, non-reciprocal transport \cite{Tokura2018NonreciprocalMaterials}, superconductivity \cite{Ganesh2014TheoreticalLayer}, magnetism \cite{Jiang2021UnconventionalKV3Sb5} and molecular design \cite{Shen2020SupramolecularFunctions}. In electronic systems, chirality can manifest in both real-space lattice configurations \cite{Zhu2024CreatingDimensions} and momentum-space electronic structures (gyrotropic order \cite{Xu2020SpontaneousDichalcogenide}), leading to chirality controlled optical activity \cite{Purcell-Milton2018InductionNanostructures} and chiral phonons \cite{Chen2018ChiralMaterials, Zhu2018ObservationPhonons}. Two-dimensional (2D) transition metal dichalcogenides (TMDs) provide an interesting platform for realizing and manipulating chiral electronic states, particularly through charge density wave (CDW) transitions \cite{Ishioka2010ChiralWaves, Song2022AtomicscaleSwitching} and rich phase diagrams \cite{Sipos2008Mott1T-TaS2, Zhao2023Spectroscopic1T-TaS2, Gao2021Chiral-TaS2}. 

The emergence of spontaneous gyrotropic electronic order where electrons organize into a chiral macroscopic state (chiral CDW) within an originally achiral lattice represents a frontier in using chirality for fundamental studies and device applications \cite{Xu2020SpontaneousDichalcogenide}. In this context, $1T\text{-}TaS_2$ emerges as a compelling candidate for investigation due to its 2D nature of the chiral CDW. Sharing the layered dichalcogenide structure and exhibiting a rich phase diagram of commensurate and incommensurate CDW states, \textit{1T}-TaS$_2$ stands out as a prototypical system where strong electron-electron correlations and electron-phonon coupling \cite{Wang2019LatticeTransitions, Singh2022Latticedriven2} drive the formation of a (nearly) commensurate CDW state featuring Star-of-David (SoD) clusters \cite{Park2019EmergentWave}, hosting planar chirality or ferro rotational order \cite{Yang2022VisualizationWaves, Singh2022Latticedriven2, Zhao2023Spectroscopic1T-TaS2, Liu2023ElectricalCrystals}. 

The prevailing model treats this as a bistable system of macroscopic left- and right-handed domains, tunable by electric fields \cite{Liu2023ElectricalCrystals}, thermal cycling \cite{Qi2024TemperatureCommensuralibity}, ultra fast pulses \cite{Zong2018UltrafastWave} and strain \cite{Qi2024InPlaneStress}. However, an in-plane electric field cannot couple linearly to the planar chirality or ferro rotational order, due to symmetry violations, yet controlled non-volatile switching has been demonstrated \cite{Liu2023ElectricalCrystals}, which is important for device applications. Isothermal chiral inversion enabled by topological defects \cite{Aishwarya2024MeltingUTe2, Cheng2024UltrafastWave, Orenstein2025DynamicalTransition} in the CDW order parameter \cite{Zong2018UltrafastWave} may underpin electric-field-driven control of planar chirality.

In this work, we use holographic analysis of scanning tunneling microscopy (STM) data \cite{Pasztor2019HolographicParameter} to image and decompose the nearly commensurate (NC)-CDW phase in \textit{1T}-TaS$_2$ single crystal surface, reconstructing the full CDW order parameter $\Delta$ and directly map the CDW amplitude and phase, which reveal phase patterns densely populated with topological defects. We classify two distinct classes of topological defects classified by homotopy groups \cite{Skogvoll2023UnifiedExcitations}: discrete chiral domains described by the homotopy group $\pi_0(\mathbb{Z}_2)$ and continuous phase defects; solitons and vortices, characterized by the homotopy group $\pi_1(S^1)$. To quantify the connection between topology and chirality, or ferro rotational order \cite{Liu2023ElectricalCrystals} we introduce a gauge-invariant chiral order parameter, $\chi(\mathbf{r}) = \mathrm{Im}(\Delta_1^* \Delta_2 + \Delta_2^* \Delta_3 + \Delta_3^* \Delta_1)$, which captures the mirror-symmetry breaking induced by SoD cluster rotation, based on the CDW phase winding. Our measurements reveal that the NC-CDW fragments into a disordered mosaic of chiral domains, forming a chiral continuum without long-range order, when rich in phase defects. By comparing regions with varying structural defect densities, we uncover a coupling between the two topological orders: a defect/strain-mediated mechanism by which $\pi_1$ phase defects, such as vortices, can pin in regions of high shear stress. Annihilation of vortex-antivortex pairs can lead to large discommensurations or soliton networks in the CDW phase.  Our proposal for chiral control by in-plane electric fields involves a soliton, serving as a precise, mobile switch to control planar chirality. This mechanism leverages the soliton as a mobile melted state of the CDW order, where the charge structure locally collapses and slips by a phase discommensuration \cite{McMillan1976TheoryTransition}. This temporary collapse lowers the free energy barrier between the double well chiral state \cite{Xu2020SpontaneousDichalcogenide}. During this brief instability, the applied electric field tilts the free energy potential well, breaking the degeneracy between the two chiral domains. Our model can explain why in-plane electric field allow for ferro rotational order control, without violating symmetries, why pinning of solitons lead to threshold voltages of chiral control and non-volatile behavior\cite{Liu2023ElectricalCrystals}, and how structural disorder and CDW topology governs charge organization, especially at elevated temperatures where the CDW is rich in disorder. 

\section*{Main}

\subsection*{Holographic imaging to extract phase maps}

\begin{figure}[b!]
    \centering
    \includegraphics[scale=0.45]{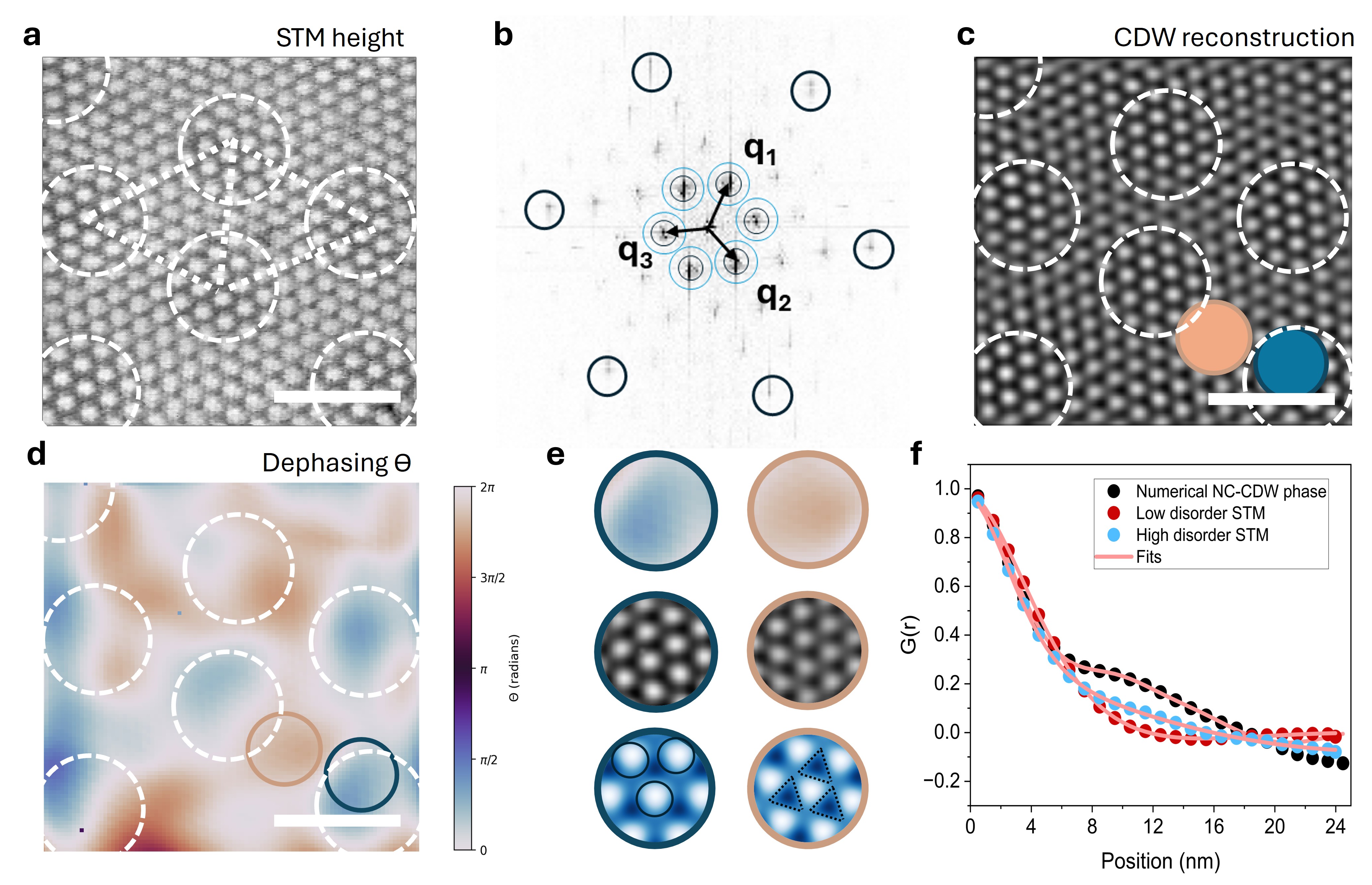} 
    \caption{
        \textbf{Holographic imaging of the CDW order parameters and the dephasing parameter $\Theta$.} 
        \textbf{(a)} STM topography map (tip bias \SI{-25}{\milli\volt}, tunnel current \SI{1}{\nano\meter}) showing the hexagonal NC-CDW phase, highlighted with the white circles, The white scale bar is equal to \SI{7}{\nano\meter}. \textbf{(b)} Fast Fourier Transform (FFT) of the STM topography. The atomic Bragg peaks are indicated with dark blue circles and the CDW peaks with light blue. The three $\vec{q_i}$ are shown of the hexagonal CDW superstructure. 
        \textbf{(c)} Real-space CDW patterns reconstruction based on the fitted phase and amplitude obtained by holographic imaging. The same structures as the STM map in (a) are overlaid for visual comparison. 
        \textbf{(d)} Corresponding spatial map of the dephasing parameter, $\Theta$. The same hexagonal CDW superstructure is indicated with the white circles. 
        \textbf{(e)} Simulated CDW patterns (lower) for different blue and orange regions $\Theta$ (upper), as indicated in (d), and compared to the reconstructed features of (c).         
        \textbf{(f)} Radial autocorrelation functions (ACF) showing decay and weak quasi-periodic structure when the disorder is low, and destruction of quasi periodic structure for large disorder. The data are fitted with stretched exponential + damped cosine, see \textbf{Supplementary \ref{Supp_sec:radial_correlation}}. 
   } 
    \label{fig:cdw_holography}
\end{figure}

We employ holographic imaging of Ref. \cite{Pasztor2019HolographicParameter} to extract spatially resolved phase and amplitude maps from STM topographic data of the NC-CDW phase in 1T-TaS$_2$, see \textbf{Supplementary section \ref{Supp_sec:holographic_imaging}} for details. A heterogeneity in the phase registry provides the foundation for our subsequent analysis of the chiral order that emerges from the interplay of the three components of the CDW q vector.
\textbf{Figure \ref{fig:cdw_holography}a} shows the STM topography (bias tip voltage \SI{-25}{\milli\volt}, tunnel current \SI{1}{\nano\ampere}), with the corresponding Fast Fourier Transform (FFT) in \textbf{Figure \ref{fig:cdw_holography}b} reveals both the atomic lattice (light blue circles) and the hexagonal CDW superstructure with three q-vectors $\vec{q}_1$, $\vec{q}_2$, and $\vec{q}_3$ (dark blue circles). Following Ref.~\cite{Pasztor2019HolographicParameter}, we decompose the CDW into complex order parameters $\Delta_i = A_i(x,y) e^{i\phi_i(x,y)}$ for each q-vector, where $A_i(x,y)$ is the local amplitude and  $\phi_i(x,y)$ the local phase. The accuracy of this decomposition is verified by reconstructing the STM image using:
$I(x, y) = \sum_{i=1}^{3} A_i(x, y) \cos(\vec{q}_i \cdot \vec{r} + \phi_i(x, y))$. The reconstructed map (\textbf{Figure~\ref{fig:cdw_holography}c}) closely reproduces the experimental topography, as indicated by the white NC-CDW domain overlay, confirming the validity of the extraction of phase and amplitude by holographic imaging.

The individual q-vector phase maps, \textbf{Supplementary figure \ref{Supp_fig:Phase_amplitudes}}, reveal significant local phase variations, with phase shifts approaching $\pi$ in certain regions. To quantify these phase relationships, we apply the dephasing parameter $\Theta = \left(\sum_{i=1}^{3} \phi_i\right) \bmod 2\pi$ \cite{Pasztor2019HolographicParameter}, which provides a direct measure of the relative phase alignment between the three CDW components. The resulting dephasing map (\textbf{Figure \ref{fig:cdw_holography}d}) reveals a complex phase structure. Regions with $\Theta \approx 0$ or $2\pi$ (reddish areas) correspond to incommensurate domains where the three CDW sublattices are poorly phase registered to the lattice, while regions with intermediate $\Theta$ values (blue/white areas) where the phases are in registry. $\Theta$ directly controls the STM resolved local CDW contrast, as demonstrated by STM simulations (\textbf{Figure~\ref{fig:cdw_holography}e}) showing that variations in $\Theta$ alone can reproduce the observed STM contrast variations from hexagonal to triangular features. 

We identify the McMillan-type structure \cite{McMillan1976TheoryTransition}, with CDW phase domains $\Theta$ near zero or $2\pi$, interrupted by narrow discommensurations walls that involve phase shifts. Notably, the local dephasing map also reveals gradual, continuous phase variations absent sharp domain boundaries, indicating that the CDW can exhibit spatially diverse phase registry separate from the lattice. The dephasing map highlights a more complex phase distribution in the NC-CDW, where domains do not arise from a singular dephasing value, as illustrated by the white circles around domains. Consequently, the NC-CDW is characterized by more than discrete commensurate domains versus incommensurate walls, distinctly exhibited by the red domains showcasing CDW domain dephasing or incommensuration. \textbf{Supplementary section \ref{Supp_sec:CDW_simulation_CCDW_NCCDW}, figure \ref{fig:Simulation}} present the dephasing maps for both ideal CCDW and NC-CDW cases, in the absence of disorder. 

\subsection*{CDW glass}
The radially averaged correlation function (ACF) applied to the dephasing maps can be used to reveal CDW phase coherence, \textbf{Figure~\ref{fig:cdw_holography}f}. To quantitatively track the effects of disorder, we analyzed three distinct cases: a numerical simulation, an STM map of a low-defect area (\textbf{Figure~\ref{fig:cdw_holography}a}), and a map from a region with high structural defect density (see \textbf{Supplementary section \ref{Supp_sec:radial_correlation}}) and \textbf{Figure \ref{fig:Dephasing_ACF}}. The ideal NC-CDW is characterized by a long short-range correlation length ($\xi_1 \approx 1.70$ nm) and a very sharp, near-Gaussian decay ($\beta_1 \approx 4.5$), reflecting a highly ordered state. It also exhibits a strong intrinsic periodicity with a wavelength of $\lambda \approx 2.23$ nm.

In the experimental low disorder limit, the oscillatory wavelength expands significantly to $\lambda \approx 3.53$ nm. This increase in the characteristic length scale suggests that in the real crystal, the NC-CDW superlattice relaxes into larger domain structures compared to the ideal simulation, likely due to pinning by static (hidden) crystal defects or macroscopic intrinsic strain. Despite these distortions, the system maintains a robust oscillatory component ($A_2 \approx 0.28$), confirming that the global hexagonal topology remains intact.

This order breaks down in the high defect density map, see \textbf{Supplementary section \ref{Supp_sec:radial_correlation}}, which provides strong quantitative evidence for a pinned CDW glass-like structure \cite{Mallayya2024BraggClustering}. Here, the decay becomes classically glassy with a stretching exponent $\beta_1 \approx 1.7$, a hallmark of disordered systems previously reported in 2D TMDs \cite{Okamoto2015ExperimentalDichalcogenide, Biljakovic1998GlassTaS3}. The oscillatory amplitude collapses ($A_2 \approx 0.07$), indicating the loss of the periodic superlattice. However, the characteristic feature size remains $\lambda \approx 3.53$ nm. This suggests that disorder does not simply reorganize the CDW into larger domains, but rather shatters the periodic lattice into disconnected, pinned domain fragments of the same intrinsic size. This fragmentation confirms that the disorder actively prevents long-range ordering, stabilizing a glass-like phase.

This glass-like structure is also noted by observing the evolution of the dephasing map between consecutive STM measurements (\textbf{Supplementary Figure \ref{fig:tip_repeated_scanning}}). After compensation for non-linear scan artifacts  (\textbf{Supplementary Methods}), we observe subtle but systematic changes: local dephasing parameter discommensurations spatially evolve. These rearrangements likely reflect a combination of thermally-activated dynamics and tip-induced perturbations. The strong STM tip's electric field (\SI{e9}{\volt\per\meter}) may interact with the CDW charge modulation discommensurations, locally lowering activation barriers for incommensurate movements, potentially accounting for the observed high depinning threshold voltage \cite{Mohammadzadeh2021RoomDevices}, and suggesting a high CDW viscosity and the possibility for plastic deformation of the CDW in charge transport within incommensurate structures \cite{Horovitz1984SolitonsSolitons}. Hence, we effectively observe with the repeated STM mapping the surface CDW discommenusration response to electrical field \cite{Zhao2023Spectroscopic1T-TaS2}. 

\subsection*{Topological defects}

\begin{figure}[t!]
    \centering
    \includegraphics[width=\textwidth]{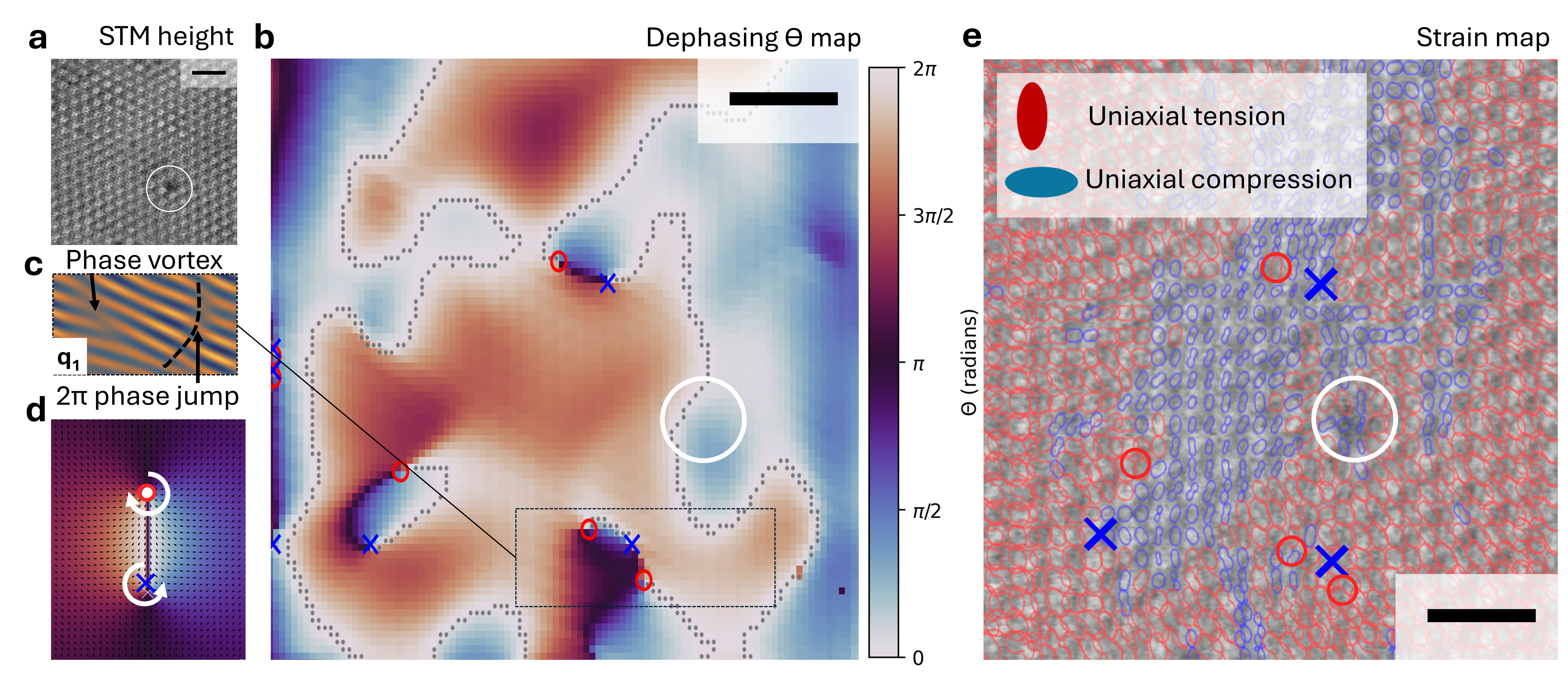}
    \caption{
        \textbf{Real-space correlation between structural defects, lattice strain, and topological phase defects.}
        (\textbf{a}) STM map, bias voltage \SI{25}{\milli\volt}, \SI{1}{\nano\ampere}, with a prominent structural defect highlighted by a white circle. Here, a star-of-david (SoD) is missing in the lattice. 
        (\textbf{b}) Corresponding dephasing map $\Theta$. The color scale indicates the phase from $0$ to $2\pi$. Red crosses and blue circles mark the locations of vortex and anti-vortex vortices where the phase winds by $\pm 2\pi$. The dotted gray lines trace the NC-CDW discommensuration network where $\Theta$ exhibits phase discommensurations or solitons. The black lines show $\pi$ phase slips connecting vortex pairs. 
        (\textbf{c}) Cut out of the $\vec{q_1}$ inverse FFT map showing the stereotypical y-fork \cite{Aishwarya2024MeltingUTe2, Mihailovic2019ImportanceApplications} indicating $\pm 2\pi$ phase vortices, as-well as solitons/discommensurations.
        (\textbf{d}) Simulated CDW phase map around a vortex-anti-vortex pair connected by a soliton, showing the $2\pi$ winding of the CDW phase.  (\textbf{e}) Local strain map overlaid on the real-space STM height map. The red and blue ellipsoids represent uniaxial tensile ($\epsilon > 0$) and compressive ($\epsilon < 0$) strain, respectively. The analysis reveals that topological vortices preferentially form at the interfaces between these opposing strain regions. In all images the black scale bar is equal to \SI{4}{\nano\meter}.}
    \label{fig:topology}
\end{figure}

The phase slips (discommensurations) and winding \cite{Brazovskii2021PhaseWaves, Moura2024McMillanGinzburgLandauDichalcogenides} in the $\Theta$ map are indicative of topological defects or dislocations \cite{Aishwarya2024MeltingUTe2}. To understand the interaction among these topological defects and how they relate to disorder chirality, we first analyze the topology following the unified field theory of Ref.~\cite{Skogvoll2023UnifiedExcitations}. The dephasing map $\Theta$ (\textbf{Figure \ref{fig:topology}b}), extracted from the STM map of \textbf{Figure \ref{fig:topology}a}, reveals defect structures that differ fundamentally in their topological classification. Following the framework of Refs.~\cite{Skogvoll2023UnifiedExcitations, Mermin1979TopologicalMedia}, we distinguish between point defects classified by homotopy theory and line defects characterized by a defect density field \cite{Skogvoll2023UnifiedExcitations}.

Topological vortices in the individual CDW phase fields $\phi_i$ ($i=1,2,3$) are classified using the first homotopy group, $\pi_1(S^1) \cong \mathbb{Z}$, appropriate for phase fields with order parameter space $S^1$. This classification assigns to each topological defect a conserved integer topological charge, or winding number $n \in \mathbb{Z}$, based on the way the phase encircles the defect core: $\oint \nabla \phi_i \cdot d\mathbf{l} = 2\pi n$. The conservation of this topological charge is a fundamental consequence of the $\mathbb{Z}$ structure of the homotopy group \cite{Skogvoll2023UnifiedExcitations}: vortices cannot disappear without annihilating in pairs or escaping to infinity. These topological defects, which exist in one of the three CDW components, become evident in the composite dephasing map $\Theta$ as localized regions where the phase winds continuously around a core. We observe vortices ($n=+1$, counter-clockwise winding, blue) and antivortices ($n=-1$, clockwise winding, red) exhibiting smooth, continuous phase rotation, as indicated with the blue and red markers in \textbf{Figure \ref{fig:topology}b}, and simulated in \textbf{Figure \ref{fig:topology}d}.

The dephasing map $\Theta$ reveals two distinct classes of line features that differ fundamentally in their physical origin. The gray dotted lines (\textbf{Figure \ref{fig:topology}b}) trace $2\pi$ phase wraps in $\Theta$ that correspond to the discommensuration network characteristic of the NC-CDW phase. At these domain walls, each of the three CDW order parameter components ($\phi_1, \phi_2, \phi_3$) undergoes a phase slip to accommodate the lattice incommensurability. Since $\Theta = (\sum_n \phi_n) \bmod 2\pi$, the simultaneous shifting of three CDW phase components manifests as a $2\pi$ phase wrap jumps in the summed field, (\textbf{Figure \ref{fig:topology}c}), causing $\Theta$ to return to zero at the domain wall centers. Note that the discommensuration value is close to $\pi$, as expected for the NC-CDW phase, but can vary indicating that this is not purerly commensurate lattice matching \cite{Pasztor2019HolographicParameter}. This discommensuration network is an intrinsic structural feature of the NC phase, topologically required to reconcile the CDW periodicity with the underlying lattice, but likely subjected to strain and defect mediated symmetry breaking. 

In contrast, the black lines connecting vortex-anti-vortex apairs indicate $\pi$ phase slips that represent energy minimization connecting the vortex pairs. These $\pi$ slips are not classified by $\pi_1(S^1)$; while vortices are characterized by drawing a 1-dimensional loop around them, one cannot draw a small loop around a 1-dimensional line defect in 2D to extract a winding number \cite{Skogvoll2023UnifiedExcitations}. Instead, we characterize them using the defect density field \cite{Skogvoll2023UnifiedExcitations}. This field captures both topological point defects and non-topological line excitations where the CDW phase and/or amplitude vary significantly, as supported by the CDW amplitude order maps given in \textbf{Supplementary figure \ref{fig:Supp_amplitude_solitons}} which show suppression of the amplitude at these line defects.

The coexistence of both the intrinsic NC discommensuration network; $2\pi$ wraps in $\Theta$ and vortex-pair induced $\pi$ phase slips suggests a rich energy landscape. The $<\pi$ slips likely represent partial discommensurations where only one or two of the three CDW components undergo phase slips, or regions where the NC matrix structure is disrupted by material disorder. In the framework of Ref.~\cite{Skogvoll2023UnifiedExcitations}, such phase slips often appear as precursory patterns or nucleation sites for topological defect formation. We denote them as solitons. A particularly interesting observation is that vortex-anti-vortex pairs can be connected by line defects: connections through the NC discommensuration network represent pathways where the phase navigates through the matrix with (partial) amplitude suppression, see \textbf{Supplementary figure \ref{fig:Supp_amplitude_solitons}}, while $\pi$ connections configurations where complete amplitude suppression occurs along the connecting line, \textbf{Supplementary figure \ref{fig:Supp_amplitude_solitons}}. This phenomenology suggests complex interactions between point defects, the NC matrix structure, and disorder-induced line defects that are influenced by strain fields. Furthermore, the presence of $\pi$ slips hints at possible fractionalization or disruption of the NC discommensuration network; disorder can locally destabilize the $2\pi/3$ fractional soliton structure, creating regions where incomplete phase slips occur, analogous to dislocation splitting in crystal plasticity and the dissociation of discommensurations in active matter systems \cite{Skogvoll2023UnifiedExcitations}.

To provide support for CDW energetics driving soliton localization and understand the complex, fibrous networks formed by the solitons, we performed a quantitative morphological analysis of their boundaries, \textbf{Supplementary section \ref{Supp_sec:fractality_model}}. We find that the soliton network exhibits a fractal geometry, see \textbf{Supplementary figure \ref{fig:supp_fractal}}. In typical regions, perimeter-area scaling analysis reveals a moderately fractal boundary with a dimension of $D_b \approx 1.20$. However, in the vicinity of a large structural defect, the network becomes more complex, with a space-filling fractal dimension of $D_b \approx 1.99$. This tunable fractality suggests that the soliton morphology is an emergent property governed by the competition between the solitons intrinsic line tension and the pinning potential from material defects. To test this hypothesis, we developed a phenomenological dynamics model, \textbf{Supplementary section \ref{Supp_sec:fractality_model}} and \textbf{Supplementary figure \ref{fig:supp_fractal}}, that simulates a flexible soliton wall in a disordered energy landscape. The model successfully reproduces our experimental findings: it not only yields a fractal dimension ($D \approx 1.12$) in excellent agreement with the discommensuration network but also demonstrates that the dimension is tunable. This agreement provides compelling evidence that the observed fractal self-organization of discommensuration is driven by material disorder and pinning \cite{Lee2023MobileLayer}.

While topology ensures stability for defects, energetics likely dictates preferred nucleation and pinning sites. The strain map (\textbf{Figure \ref{fig:topology}d}) reveals that the structural vacancy creates regions characterized by high compressive and tensile lattice stress (see \textbf{Supplementary Methods} for strain extraction methodology). These regions feature an uneven local strain field with spatially adjacent tensile and compressive areas. The vortex core, an inherently high-energy zone, requires the CDW order parameter's magnitude to nullify. To minimize total free energy, we propose the CDW nucleates or pins vortex cores at high-stress boundaries, reducing the energetic cost of co-existing topological defects and strain fields. This illustrates the interaction between topological defects (vortex protected by $\pi_1$) and non-topological excitations (strain field with continuous relaxation).

\subsection*{Chiral order parameter}

Having established the full CDW order of the NC-CDW phase and the hosting of topological defects, we turn to the chirality of the SoD rotation or ferro-rotational order \cite{Liu2023ElectricalCrystals, Qi2024TemperatureCommensuralibity}. We diverge from the 3D chiral CDW definition \cite{Ishioka2010ChiralWaves}, i.e. for TiSe$_2$ \cite{Ishioka2010ChiralWaves}, and introduce chirality in the SoD phase winding of the 2D CDW nature of TaS$_2$. Consequently, we introduce a chiral order parameter, $\chi(\mathbf{r})$, which is gauge-invariant, reverses sign under mirror inversion, and resides solely in 2D. We constructed a parameter from the complex CDW order parameters, $\Delta_j = A_j e^{i\phi_j}$. The simplest expression satisfying the required symmetries is derived from the cyclic combination $C = \Delta_1^* \Delta_2 + \Delta_2^* \Delta_3 + \Delta_3^* \Delta_1$, in \textbf{Supplementary section \ref{Supp_sec:derivition_chiral_order}},  the derivation is given. The chiral component is the imaginary part of this term, as it is odd under a mirror reflection symmetry operation. This defines our order parameter:
\begin{equation}
\chi(\mathbf{r}) = \text{Im}(C) = A_1A_2\sin(\phi_2-\phi_1) + A_2A_3\sin(\phi_3-\phi_2) + A_3A_1\sin(\phi_1-\phi_3)
\label{eq:chi_main}
\end{equation}

\begin{figure}[h!]
    \centering
    \includegraphics[width=\textwidth]{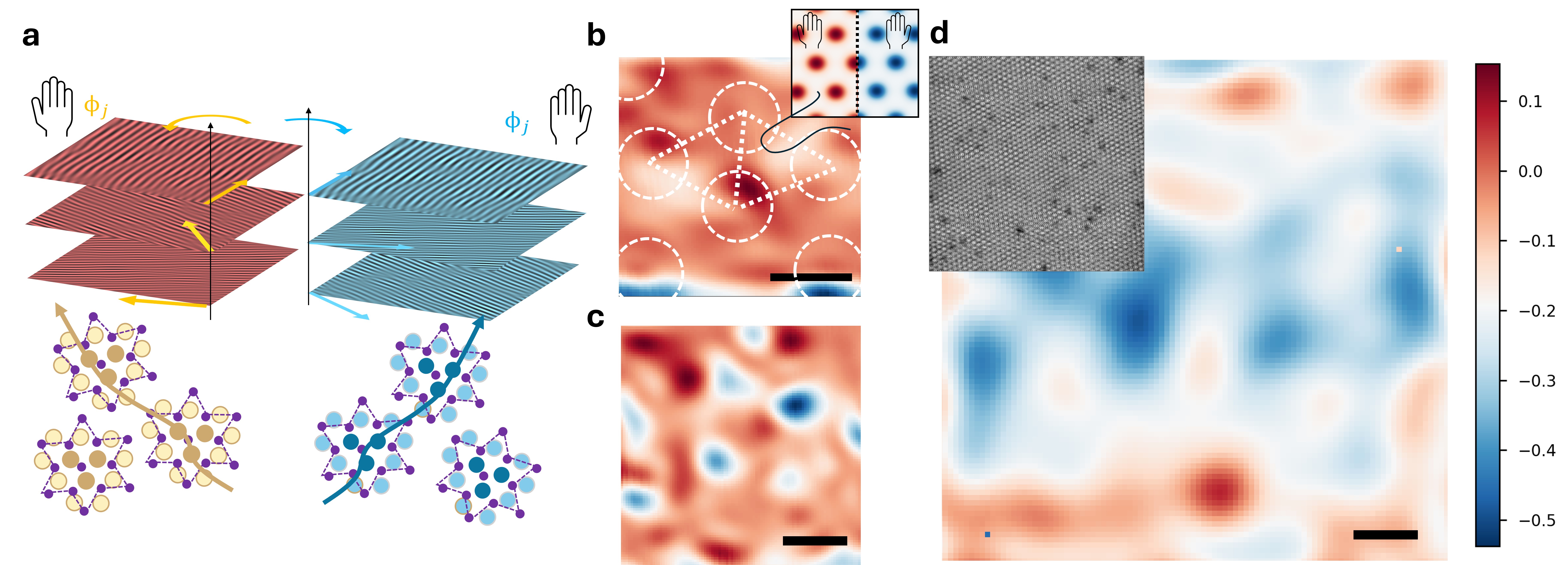}
    \caption{\textbf{The CDW phase winding chirality.} 
    (\textbf{a}) The three complex CDW order parameters $\Delta_i$ are decomposed in their respective phase maps $\phi$, the relative winding of the phases encodes the SoD rotation or planar chirality. The rotation strength is encoded in the scalar nature of $\chi$.
    (\textbf{b}) Chiral $\chi$ map of the low defect density STM map, overlaid with the NC-CDW domain structure (dashed white circles). Overall, the CDW is in one chiral state, closely matching the winding of the CDW \cite{Singh2022Latticedriven2}. The inset shows the calculated NC-CDW $\chi$ map, without any disorder, were the chirality or CDW phase winding is stronger in the domain structure. 
    (\textbf{c}) $\chi$ map of the high defect density STM map, which hosts high density of discommensurations and lead to a fragmented chiral state. 
    (\textbf{d}) Large area STM height map (\SI{50}{}x\SI{50}{\nano\meter}) with high defect density and $\chi$ map, highlighting the fragmented chiral state and local strong chiral nature of the CDW phase winding. The black scale bar in all images is equal to \SI{5}{\nano\meter}.}
    \label{fig:chirality_model}
\end{figure}

This quantity is non-zero only when the CDW phase wind in a consistent direction, providing a direct, local measure of planar SoD chirality. Crucially, $\chi$ is odd under both mirror inversion ($\mathcal{M}$) and time-reversal ($\mathcal{T}$) symmetry, see symmetry analysis in \textbf{Supplementary section \ref{Supp_sec:symmetry_chiral_order}}. However, because we measure topological structures with the STM, time-reversal symmetry becomes even again.  By mapping $\chi$, we directly image the local electronic chirality of the CDW state. In the NC-CDW phase, this electronic chirality can be interpreted as being coupled to the structural ferro-rotational order \cite{Liu2023ElectricalCrystals} ($\Omega$), such that domains of $\chi > 0$ correspond to regions where the CDW SoD are rotated by $+\phi$, and domains of $\chi < 0$ correspond to rotation by $-\phi$. While both order parameters $\Omega$ and $\chi$ break mirror symmetry, the electronic parameter $\chi$ is a P-odd pseudoscalar, whereas the $\Omega$ transforms as a P-even axial vector. However, on the surface, $\Omega$ should also be P-odd, see discussion in \textbf{Supplementary section \ref{Supp_sec:symmetry_chiral_order}}.
While $\Omega$ treats the CDW domains as two handedness of the SoD rotation, STM experiments of the NC-CDW \cite{Singh2022Latticedriven2} show swirling chiral CDW gradients. Therefore, we employ $\chi$ because it inherently couples the electronic chirality to the local order parameters phase and amplitude with a scalar nature to encode the strength of phase winding, \textbf{Figure \ref{fig:chirality_model}a}. To provide a quantitative and intuitive framework for visualizing chirality, we introduce an amplitude-weighted phase triangle model, see \textbf{Supplementary section \ref{Supp_sec:phase_triangle_chiral_order}} for  derivation. A positive area corresponds to a counter-clockwise orientation of the phase vectors, while a negative area indicates a clockwise orientation. An achiral state, where the vertices are collinear, has zero area. This model captures how both phase relationships and amplitude imbalances contribute to net chirality which relates directly to the STM CDW contrast.

In surface regions with minimal structural disruption, evident in the STM map of \textbf{Figure \ref{fig:cdw_holography}a}, the NC-CDW domains exhibit stronger phase winding and intra-domain chirality \cite{Singh2022Latticedriven2}, \textbf{Figure \ref{fig:chirality_model}b}. In areas abundant with structural defects that locally disturb the hexagonal order, a CDW glass forms. This is accompanied with a fragmentation of the chiral texture, \textbf{Figure \ref{fig:chirality_model}c, d}, indicating that discommensurations, following local CDW perturbation, influence the chiral organization. 

With the chiral domain structure characterized, we can now address the relationship between the two distinct topological orders that coexist within the NC-CDW state, \textbf{Supplementary figure \ref{Supp_sec:chirality_simulation_NCCDW}}. The first order governs phase defects, such as vortices and discommensurations, which are classified by $\pi_1(S^1)$, phase, and represent the windings of the overall CDW phase. The second governs the chiral domains, which are classified by $\pi_0(\mathbb{Z}_2)$ and depend only on the relative phases between the three CDW components.  Critically, we observe that the discommensurations network and the chiral domain walls do not spatially coincide, see \textbf{Supplementary figure \ref{topology_decoupling}}. This provides direct experimental evidence for a fundamental decoupling between the $\pi_1$ (phase) and $\pi_0$ (chirality) topologies. . 

\section*{Discussion}

\begin{figure}[]
    \centering
    \includegraphics[scale=0.6]{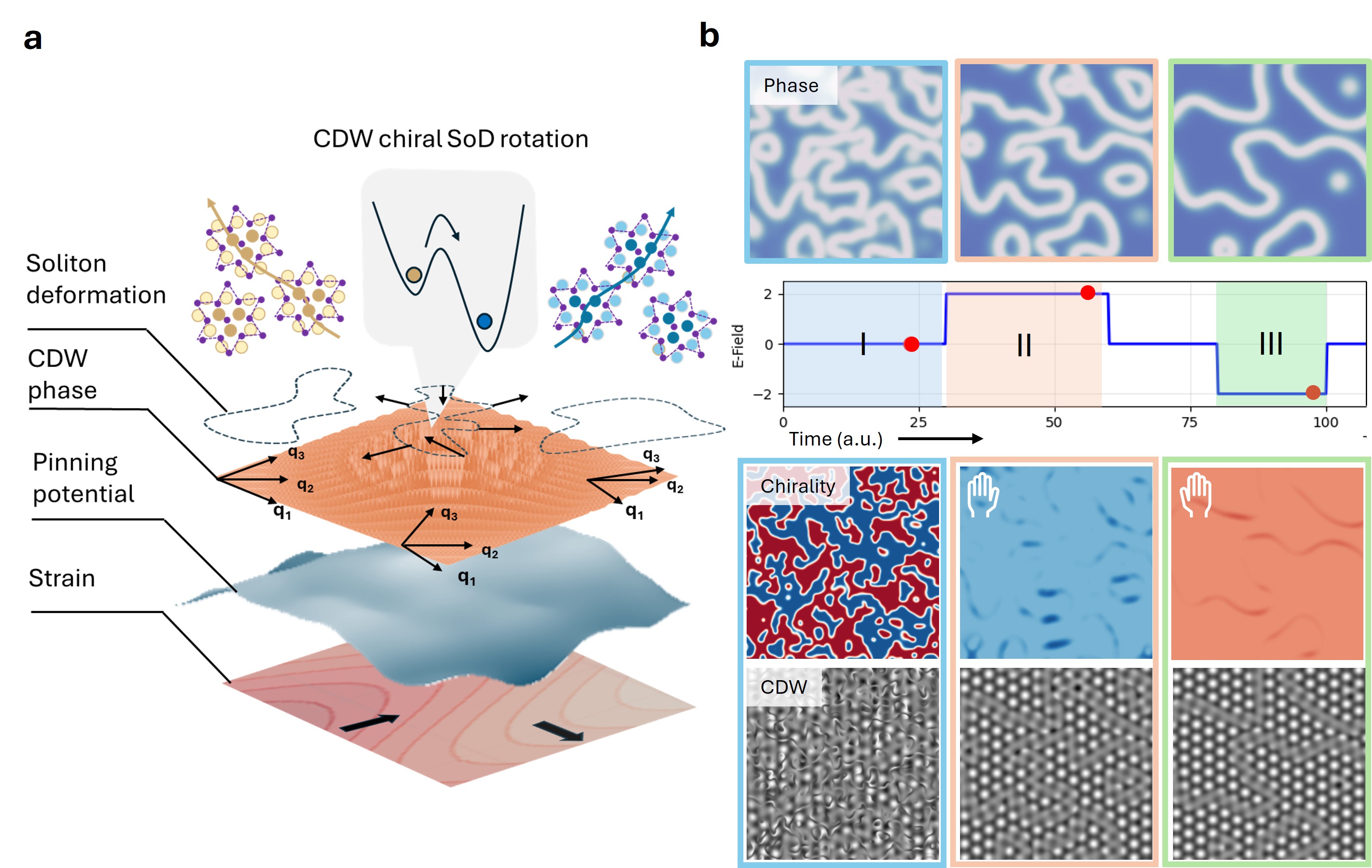} 
    \caption{\textbf{The proposed and modeled mechanism for chiral inversion.} (\textbf{a}) The process is schematically illustrated, where topological defects are nucleated and perturbed by strain \cite{Qi2024InPlaneStress}, temperature \cite{Qi2024InPlaneStress, Liu2023ElectricalCrystals}, or an $\mathbf{E}$-field \cite{Liu2023ElectricalCrystals}, and can be immobilized by local defects within the pinning potential landscape. Defect nucleation and their subsequent annihilation contribute to the formation of residual solitonic structures. We propose that the temporary distortion and evolution of soliton organization, for example by an $\mathbf{E}$-field, allow for the concurrent rearrangement of the three $\vec{q}$ vector wavefronts, altering their chiral CDW phase winding direction or SoD rotation. (\textbf{b}) A Ginzburg-Landau model simulation (\textbf{Supplementary \ref{Supp_sec:chiral_inversion_model} and Movie S1 for details}) of the formation of solitons following the nucleation and annihilation of vortex pairs (stage I), during which the CDW fragments into two chiral states (depicted as red and blue domains). When an $\mathbf{E}$-field pulse with opposite polarity is applied (phase II, III), it reverses the chirality through the movement of solitons, causing the CDW order parameter $\Delta$ to locally melt, allowing for traversing the chiral double well potential. The images, CDW phase, chirality and CDW order, are from frames (red dot) of the phases as indicated, see \textbf{Supplementary Movie S1}. 
   } 
    \label{fig:discussion}
\end{figure}

In our study we identify chiral domains, classified by the zeroth homotopy group $\pi_0(\mathbb{Z}_2)$ (representing two discrete ground states), alongside a network of phase defects (vortices) classified by the first homotopy group $\pi_1(S^1)$ (representing continuous phase windings). These two orders are topologically independent. A vortex, as a $\pi_1(S^1)$ defect, cannot directly convert into a $\pi_0(\mathbb{Z}_2)$ chiral wall; a direct topological conversion is forbidden. We propose that the system bypasses this constraint via an indirect, dynamic coupling via solitons or phase discommensurations. A perturbation of the CDW order parameter $\Delta$, i.e., by strain \cite{Qi2024InPlaneStress}, thermal cycling \cite{Qi2024InPlaneStress, Liu2023ElectricalCrystals} or light pulses \cite{Zong2018UltrafastWave} nucleates a large number of phase vortex pairs, where some can be pinned by structural defects; \textbf{Figure  \ref{fig:discussion}a}. Overall, these pairs annihilate and their connecting soliton form large  network structures. These solitons, frozen by the dense pinning landscape, lead to a glass-like CDW structure of the NC-CDW. This dense glass of small loops subsequently coarsens over time, merging to minimize total line energy, which accounts for the large, fractal networks we observe. This framework explains why regions of high structural disorder are conspicuously free of isolated vortices and instead dominated by large, closed soliton loops. The soliton network's arrangement is not governed by the bulk strain field only, but rather by the microscopic pinning landscape created by local strain variations and impurities. This aligns with x-ray microdiffraction \cite{Bellec2020EssentialFields}, which show that defect dynamics are dominated by strong pinning, at least for 1D CDW systems. In this context, the CDW acts as a viscous medium \cite{Delacretaz2017TheoryStates}, where subtle soliton movement \cite{Lee2023MobileLayer} accounts for the discommensurations network deformation observed between STM scans, where a strong electric field is required to excite mobility above the pinning threshold \cite{Brown2023CurrentFilms}.

A Ginzburg-Landau model was developed to establish a possible mechanism by which an external $\mathbf{E}$-field controls the chiral switching, \textbf{Figure \ref{fig:discussion}b}, and details are given in \textbf{Supplementary section \ref{Supp_sec:chiral_inversion_model}}, which can explain the experimental work on electric field control of 1T-TaS$_2$ ferro rotational order \cite{Liu2023ElectricalCrystals}. First, soliton structures are created by nucleation and annihilation of vortex pairs, \textbf{Fig \ref{fig:discussion}b, phase I}. We define the soliton as a mobile, topological catalyst that gates the process: the $\mathbf{E}$-field first drives the charged soliton  across the CDW landscape, \textbf{Fig \ref{fig:discussion}b, phase II, III}. As the soliton core passes, it induces a local collapse or melting of the CDW amplitude ($A_i$), thereby removing the energy barrier that locks the commensurate phase structure into its current chiral state ($+\chi$ or $- \chi$). In this brief, softened CDW state window, the $\mathbf{E}$-field asserts directional control by applying a bias to the Ginzburg-Landau free energy potential \cite{Xu2020SpontaneousDichalcogenide} of the chiral order parameter ($\chi$). This bias forces the local symmetry to flip to the energetically preferred state dictated by the field's sign, following a canted double well potential. After the pulse has passed, a stable non-volatile chiral state is achieved. The overall result is that the $\mathbf{E}$-field controls chirality not by directly breaking the 3Q phase registry, but by utilizing the soliton's passage to momentarily melt the CDW order and thus commensurability for the chiral domain to invert along its trajectory.  Therefore, the $\mathbf{E}$-field (in-plane at least) does not violate the symmetry of ferro rotational order control \cite{Liu2023ElectricalCrystals}.

This work provides the missing microscopic mechanism for the manipulation of chiral domains previously observed \cite{Zong2018UltrafastWave}. It was demonstrated that a femtosecond light pulse can switch the global chiral state by creating a high concentration of topological defects. We propose that the photo-induced transient ``melted" phase they observe is the non-equilibrium equivalent of the vortex nucleating and annihilation and subsequent plastic deformation of the $\pi_1(S^1)$ defect glass we observe. This transient deformation of the CDW is what drives the system across the energy barrier to reconfigure the chiral state. 

Our application of holographic imaging to STM data of \textit{1T}-TaS$_2$ has allowed us to look beyond surface topography, deconstructing the nearly commensurate CDW into its constituent complex order parameters $\Delta_i$ and identifying a distinct "CDW glass" phase. By mapping the local dephasing parameter $\Theta$,  we reveal that structural disorder does not merely perturb the lattice but stabilizes a dense, pinned network of topological defects. This analysis uncovers a fundamental topological decoupling within the material. This distinction resolves the paradox of how an in-plane electric field, which cannot linearly couple to the planar chirality due to symmetry constraints, achieves deterministic control over the ferro-rotational order. Ultimately, by linking the G-L theory of the CDW to the microscopic interplay between structural pinning and topological textures, our work improves our understanding of control of planar chirality and potentially enables future investigations involving spin \cite{Menichetti2025ChiralityinducedDichalcogenides} and chiral phonons \cite{Pan2024StraininducedWS2}, in technological applications, for example, memory.


\section*{Acknowledgments}
We express thanks to ing. Wijnand Dijkstra for technical assistance with the STM. 
\paragraph*{Funding:}
M.V. and C.F.J. Flipse gratefully acknowledge financial support from NWO MagCat OCENW.M.21.104. M.L. acknowledge financial support from NWO QuMat.  This research was conducted within the Eindhoven University of Technology, Department of Applied Physics and Science Education. 
\paragraph*{Author contributions:}
M.V. conducted the experiments, wrote the python codes and analyzed the data. M.L and C.F.J. Flipse supervised the research. All authors contributed to the interpretation and manuscript writing. 
\paragraph*{Competing interests:}
There are no competing interests to declare.
\paragraph*{Data and materials availability:}
The python code used in this work is available on Github.

\newpage


\subsection*{Supplementary materials}
Materials and Methods\\
Supplementary Text S1 to S12\\
Figs. S1 to S12\\
References \textit{(7-\arabic{enumiv})}\\ 


\newpage


\renewcommand{\thefigure}{S\arabic{figure}}
\renewcommand{\thetable}{S\arabic{table}}
\renewcommand{\theequation}{S\arabic{equation}}
\renewcommand{\thepage}{S\arabic{page}}
\setcounter{figure}{0}
\setcounter{table}{0}
\setcounter{equation}{0}
\setcounter{page}{1} 


\begin{center}
\section*{Supplementary Materials for\\ \scititle}

Michael Verhage,
Martin lee,
Kees Flipse$^\ast$\\ 
\small$^\ast$Corresponding author. Email: c.f.j.flipse@tue.nl\\
\end{center}

\subsubsection*{This PDF file includes:}
Materials and Methods\\
Supplementary Text S1 to S10\\
Figures S1 to S10\\
Caption for Movie S1\\

\subsubsection*{Other Supplementary Materials for this manuscript:}
Movie S1

\setcounter{section}{0}
\renewcommand{\thesection}{S\arabic{section}}

\newpage
\section{Holographic imaging} \label{Supp_sec:holographic_imaging}

The holographic imaging method decomposes the STM topography into its constituent CDW components, we follow the work of Ref. \cite{Pasztor2019HolographicParameter}. 
The complex order parameter for each component, $\Psi_j(\mathbf{r}) = A_j(\mathbf{r}) e^{i\phi_j(\mathbf{r})}$, is extracted via Fourier analysis. This involves:
\begin{enumerate}
    \item Identifying the primary CDW wave vectors, $\mathbf{q}_j$, from peaks in the FFT of the STM image.
    \item Applying a circular mask in Fourier space around each $(\mathbf{q}_j, -\mathbf{q}_j)$ pair to isolate the signal bandwidth. The mask radius, $R_{\text{mask}}$, is a key parameter balancing noise suppression and spatial resolution. We made sure to choose the radius as large as possible, to capture local phase distortion, but made sure not to overlap with the neighboring peak radii. 
    \item Performing an inverse FFT on the masked data to reconstruct the complex field $\Psi_j(\mathbf{r})$.
\end{enumerate}
\begin{figure}[b!]
    \centering
    \includegraphics[width=0.7\textwidth]{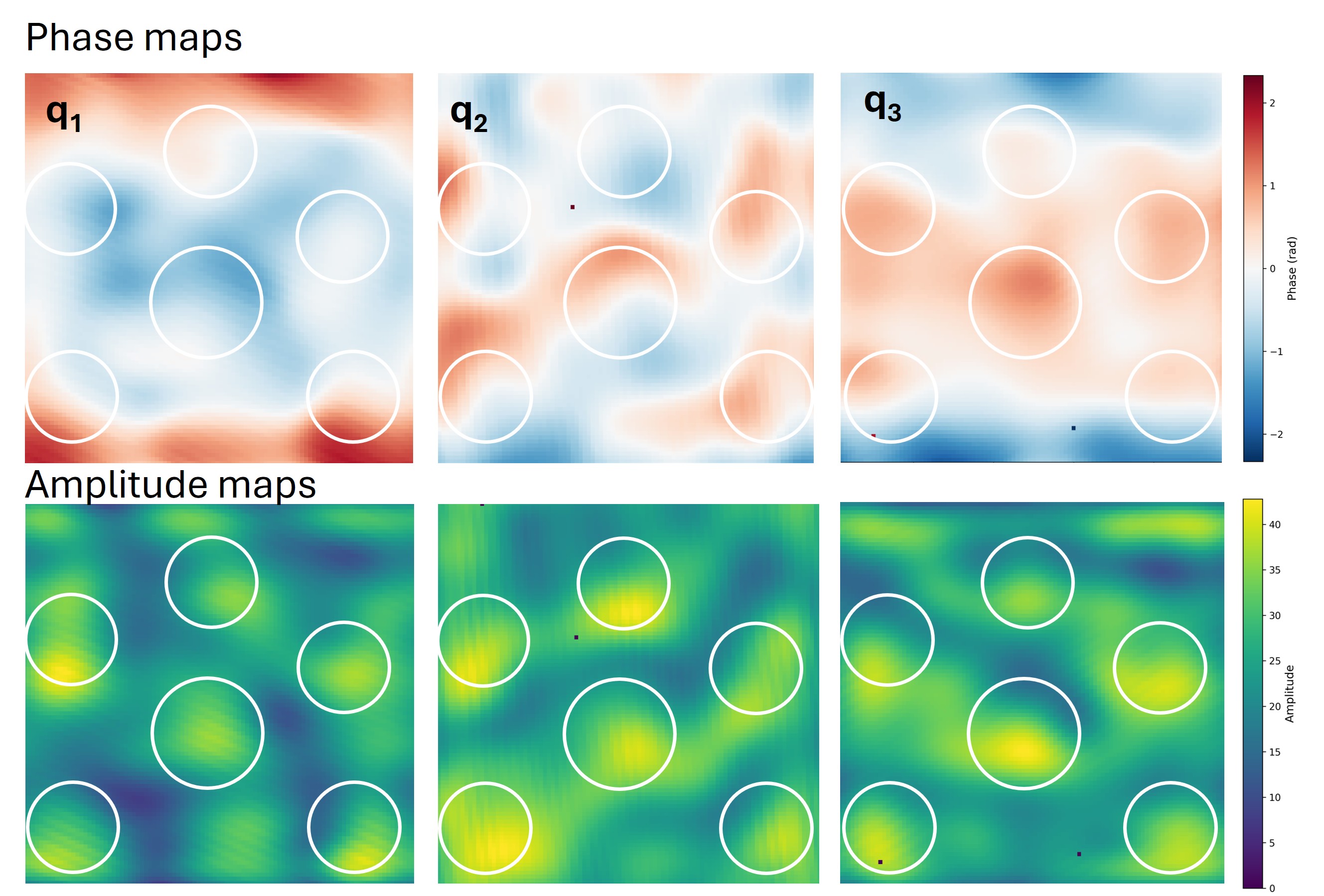}
    \caption{
        \textbf{Extracted phase and amplitude maps of the $q_i$ vectors.} The white circles indicate NC-CDW domains. Note the higher amplitude for domains, which relates to the higher intensity in the STM maps in resolving the NC-CDW CDW contrast.
    }
    \label{Supp_fig:Phase_amplitudes}
\end{figure}
To obtain detailed spatial maps of amplitude and phase, we employ a sliding window fitting algorithm \cite{Pasztor2019HolographicParameter}.
\begin{itemize}
    \item \textbf{Local fitting:} Within each analysis window of size $W \times W$ (typically $W=32$ pixels) with a step size $S$ (typically $S=W/2$), we fit the model:
    \begin{equation}
    f(x,y) = A \cos(q_x \cdot x + q_y \cdot y + \phi)
    \end{equation}
    using the Levenberg-Marquardt algorithm. Initial guesses for the fit are derived from the global phase maps to ensure robust convergence.
    
    \item \textbf{Quality estimation:} The goodness-of-fit in each window is quantified by the coefficient of determination, $R^2$. Fits with $R^2 < 0.9$ are typically excluded from further analysis.
    \begin{equation}
    R^2 = 1 - \frac{\sum_i (I_i - f_i)^2}{\sum_i (I_i - \bar{I})^2}
    \end{equation}
\end{itemize}

\subsection*{Phase flattening and the composite dephasing map}

The extracted phase maps $\phi_j(\mathbf{r})$ from the inverse FFT procedure contain both intrinsic CDW phase modulations and extrinsic contributions arising from experimental artifacts and sample inhomogeneities. In particular, the finite width of FFT peaks observed as broadened distributions rather than sharp delta functions in Fourier space, indicates spatial variation of the CDW wavevector $\mathbf{q}_j(\mathbf{r}) = \mathbf{q}_0 + \delta\mathbf{q}_j(\mathbf{r})$. This wavevector variation, which potentially arises from strain gradients, disorder pinning, sample tilt, the intrinsic domain structure of the NC phase itself, and the finite resoltion of the STM images, need to be taken into account to obtain the correct q-vector. 

\textbf{Origin of spurious phase gradients:} When the CDW wavevector varies spatially, the complex order parameter takes the form:

\begin{equation}
\Psi_j(\mathbf{r}) = A_j(\mathbf{r}) \exp\left[i(\mathbf{q}_0 + \delta\mathbf{q}_j(\mathbf{r})) \cdot \mathbf{r} + i\phi_{\text{true},j}(\mathbf{r})\right]
\end{equation}
The inverse FFT around the nominal wavevector $\mathbf{q}_0$ extracts a phase:
\begin{equation}
\phi_{\text{raw},j}(\mathbf{r}) = \delta\mathbf{q}_j(\mathbf{r}) \cdot \mathbf{r} + \phi_{\text{true},j}(\mathbf{r})
\end{equation}

where the first term represents a spurious contribution from wavevector variation. For typical experimental conditions with $|\delta\mathbf{q}_j|/|\mathbf{q}_0| \sim 1$-$5\%$ and fields of view of $50$-$100$ nm, this artifact can accumulate to several $2\pi$ radians, obscuring the intrinsic phase structure.

\textbf{Numerical example:} Consider a CDW with wavelength $\lambda_0 = 1$ nm (giving $q_0 = 2\pi/\lambda_0 = 6.28$ nm$^{-1}$) exhibiting a $2\%$ wavevector variation: $\delta q = 0.02 \times q_0 = 0.126$ nm$^{-1}$. Over a field of view of $50$ nm, the spurious phase accumulates to:

\begin{equation}
\phi_{\text{spurious}} = \delta q \times L = 0.126 \text{ nm}^{-1} \times 50 \text{ nm} = 6.3 \text{ rad} \approx 1.0 \times 2\pi
\end{equation}

This spurious contribution is comparable to or larger than the intrinsic phase modulations associated with discommensurations ($\sim \pi$ or $2\pi/3$), making raw phase maps difficult to interpret.

\textbf{Phase flattening procedure:} To isolate the intrinsic phase structure, we apply a flattening procedure that removes low-order polynomial backgrounds (typically linear or quadratic) from each phase map:

\begin{equation}
\phi_{\text{flat},j}(\mathbf{r}) = \phi_{\text{raw},j}(\mathbf{r}) - \left[a_j x + b_j y + c_j\right]
\end{equation}

The coefficients $(a_j, b_j, c_j)$ are determined by least-squares fitting to the raw phase map, effectively removing the dominant gradient contribution $\langle\delta\mathbf{q}_j\rangle \cdot \mathbf{r}$ while preserving local phase variations. 

\textbf{Validation through simulations:} We verified this approach using simulations of CDW phase fields with embedded topological defects (vortices). For simulated data with perfectly uniform wavevector (sharp FFT peaks), raw wrapped phases accurately recover the defect structure, and flattening offers no advantage. Note that this is simlar to the work in Ref. \cite{Pasztor2019HolographicParameter} because their FFT peaks are fitted to obtain their peak position and width. However, when we introduce realistic wavevector variations of $1$-$5\%$ to simulate experimental conditions (producing broadened FFT peaks), the raw phases become contaminated by spurious gradients, and flattening becomes essential to correctly identify defect locations and phase structures. This confirms that the necessity of flattening in experimental data arises from the finite width of FFT peaks, which reflects genuine spatial variation of the CDW wavevector, especially at \SI{300}{\kelvin} where the stability of the STM is already lower than at cryogenic temperatures. 

\textbf{Construction of the dephasing map:} After flattening, we construct the composite dephasing map by summing the three phase components:
\begin{equation}
\Theta(\mathbf{r}) = \left[\phi_{\text{flat},1}(\mathbf{r}) + \phi_{\text{flat},2}(\mathbf{r}) + \phi_{\text{flat},3}(\mathbf{r})\right] \bmod 2\pi
\end{equation}
This quantity is particularly useful for visualizing the NC-CDW domain structure. In our flattened dephasing maps, we observe two distinct phase regions: commensurate domains with $\Theta \approx 3\pi/2$ (appearing red) and matrix regions with $\Theta \approx \pi/2$ (appearing blue), separated by domain walls where $\Theta$ undergoes phase slips of approximately $2\pi$.  While these transitions may appear as $2\pi$ wraps/jumps, under the mod $2\pi$ operation, the physical phase slip is varied, and smaller than $2\pi$. 


The accuracy of the extracted parameters is validated by reconstructing the CDW signal and comparing it to the original STM data. For well-ordered regions, this method typically achieves high fidelity, with $R^2 > 0.8$. The analysis involves a fundamental trade-off between spatial resolution and fitting stability, primarily controlled by the window size $W$ and the Fourier mask radius $R_{\text{mask}}$. Similar to the claims of Ref. \cite{Pasztor2019HolographicParameter}, window size does not lead to structural changes in the extracted phase maps, only the resolution. 

\newpage
\section{STM map simulation of the CCDW and NC-CDW and extracted dephasing map}
\label{Supp_sec:CDW_simulation_CCDW_NCCDW}

\begin{figure}[!htp]
    \vspace*{\fill}
    \centering
    \includegraphics[width=0.9\textwidth]{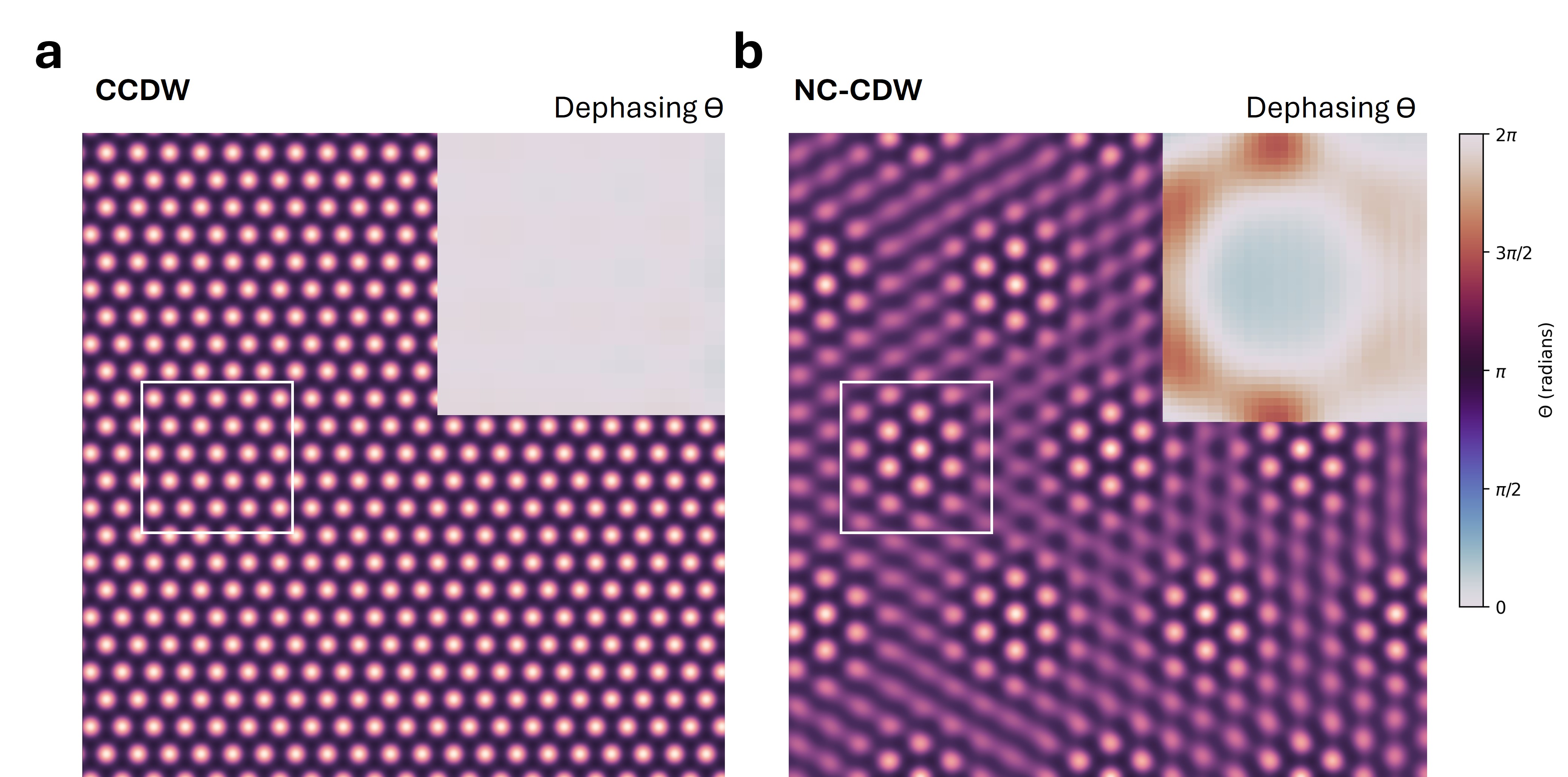}
    \caption{
        \textbf{STM simulation of the commensurate (C) CDW phase (a, i.e. at low temperature) and NC-CDW phase (b, room temperature).} Note the clear dephasing contrast in the domain regions, which are evident in the inset dephasing map of the white box, showing a $\theta$ around $3 \pi/2$, while the domains are commensurate with a $\theta$ close to $0$ or $2\pi$. The NC-CDW shows a discommensuration, white areas of phase jump, surrounding the commensurate domain.}
    \label{fig:Simulation}
    \vspace*{\fill}
\end{figure}

\newpage
\section{Radial correlation analysis of the NC-CDW phase}
\label{Supp_sec:radial_correlation}

To quantitatively analyze the spatial correlations in the numerical, low defect density, and structural disordered NC-CDW dephasing maps, the radially averaged correlation function, $G(r)$, was fitted to the dephasing maps. A model combining a stretched exponential with a damped cosine function was fitted. This model effectively captures both the short-range decay of phase coherence and the long-range oscillatory behavior characteristic of the hexagonal NC-CDW domains. The fitting function is given by:

\begin{equation}
    G(r) = A_1 \exp\left(-\left(\frac{r}{\xi_1}\right)^{\beta_1}\right) + A_2 \exp\left(-\frac{r}{\xi_2}\right) \cos(k r + \phi)
    \label{eq:fit_model}
\end{equation}

where the first term describes the primary decay of correlation with a characteristic length $\xi_1$ and a stretching exponent $\beta_1$, while the second term captures oscillations with a wavelength $\lambda = 2\pi/k$ that decay over a length scale $\xi_2$.

\begin{figure}[!htp]
    \vspace*{\fill}
    \centering
    \includegraphics[width=0.9\textwidth]{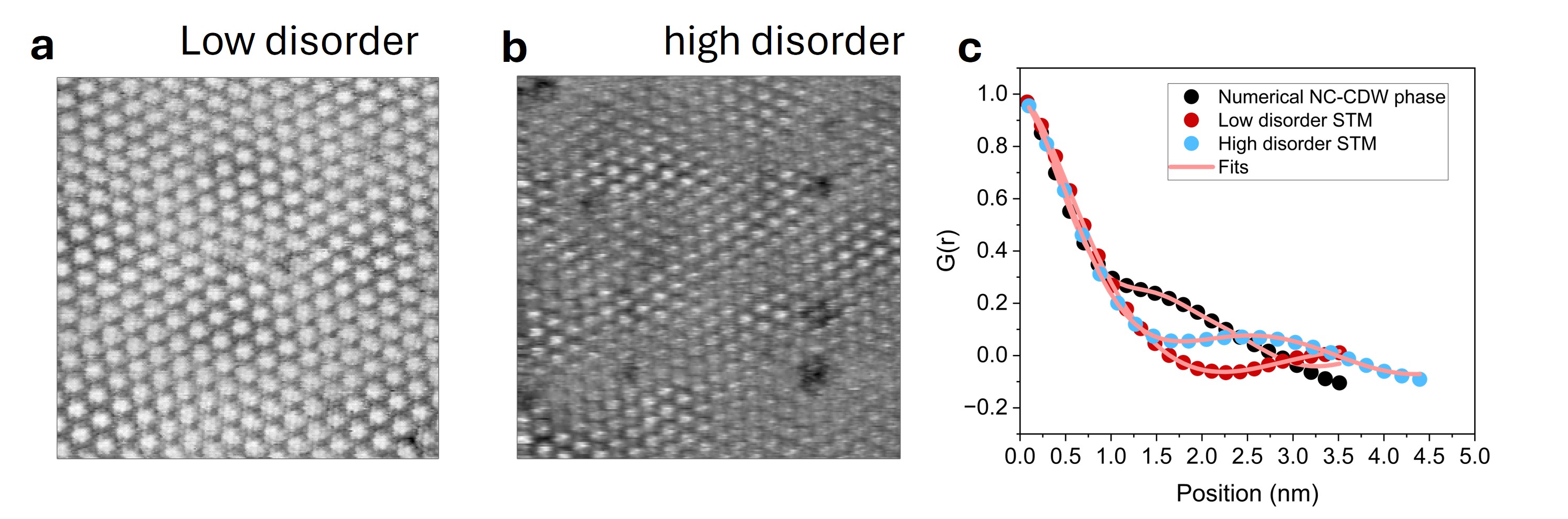}
    \caption{
        \textbf{STM map taken at low and high structural disorder.} (\textbf{a}) STM map \SI{20}{}x\SI{20}{\nano\meter} (-25mV, 1nA) with low structural disorder. (\textbf{b}) STM map \SI{25}{}x\SI{25}{\nano\meter} (-25mV, 1nA) of a high defect density rich surface area. Both maps are converted to their dephasing maps and used to calculate the radial correlation function (\textbf{c}).} 
    \label{fig:Dephasing_ACF}
    \vspace*{\fill}
\end{figure}

The best-fit parameters for the numerical NC-CDW, the low local structural disorder case, and the high structural defect density case \textbf{Figure \ref{fig:Dephasing_ACF}}, are summarized in \textbf{Table \ref{tab:fit_params}}. The results reveal a systematic breakdown of the NC-CDW's long-range order as structural disorder is introduced.

\begin{table}[h!]
    \centering
    \caption{Fit parameters for Eq.~\ref{eq:fit_model} of the correlation function for the numerical, low disorder, and disordered NC-CDW STM dephasing maps. Experimental length scales are converted to nanometers based on scan sizes of 20 nm (Low disorder) and 25 nm (Disordered). Numerical values are in pixels.}
    \label{tab:fit_params}
    \begin{tabular}{@{}lccc@{}}
        \toprule
        \textbf{Parameter} & \textbf{Numerical} & \textbf{Low disorder} & \textbf{Disordered} \\
         & (Simulation) & (20 nm scan) & (25 nm scan) \\
        \midrule
        $A_1$ (Short-range amp.)      & 0.550    & 0.834   & 0.992 \\
        $\xi_1$ (Short-range length)  & 21.71 px & 0.73 nm & 0.90 nm \\
        $\beta_1$ (Short-range exp.)  & 4.54     & 1.48    & 1.68  \\
        \midrule
        $A_2$ (Oscillatory amp.)      & 0.447    & 0.276   & 0.073 \\
        $\xi_2$ (Oscillatory decay)   & 17.97 px & 1.63 nm & -- \\
        $\lambda$ (Wavelength/Size)   & 28.56 px & 3.53 nm & 3.53 nm \\
        \bottomrule
    \end{tabular}
\end{table}

The analysis of the short-range correlation parameters reveals a shift in the dominant mechanism of decay. In the ideal numerical case, the correlation is balanced between oscillatory and exponential terms ($A_1 \approx A_2$). However, in the experimental maps, the short-range exponential term becomes dominant. In the low disorder case, $A_1$ accounts for 83\% of the correlation magnitude, rising to over 99\% in the disordered case. While the short-range correlation length $\xi_1$ remains comparable between the two experimental regimes ($0.73 - 0.90$ nm), the dramatic increase in $A_1$ in the disordered sample signifies that phase coherence is effectively confined to this short length scale. Additionally, the decay exponent $\beta_1$ drops significantly from the numerical value of 4.54 to values below 2 in the experiments (1.48 and 1.68).

The most profound structural change is observed in the long-range oscillatory behavior. In the low disorder sample, a clear oscillatory component persists ($A_2 = 0.276$) with a defined wavelength of $\lambda = 3.53$ nm. This length scale corresponds to the periodic spacing of the NC-CDW domain superstructure. In contrast, for the disordered sample, the oscillatory amplitude collapses to near-noise levels ($A_2 = 0.073$), indicating the complete destruction of the periodic superlattice. Remarkably, the fit extracts an identical characteristic length scale of $\lambda = 3.53$ nm. In the absence of periodic order, this length no longer represents a lattice constant, but rather identifies the characteristic size of the broken, pinned domain fragments. The fitting analysis thus provides a clear physical picture: disorder does not stretch the CDW lattice; instead, it shatters the periodic superstructure into disconnected domains of similar intrinsic size, driving the system into a glassy state lacking long-range phase coherence.

\newpage

\begin{figure}[!htp]
    \vspace*{\fill}
    \includegraphics[width=1.0\textwidth]{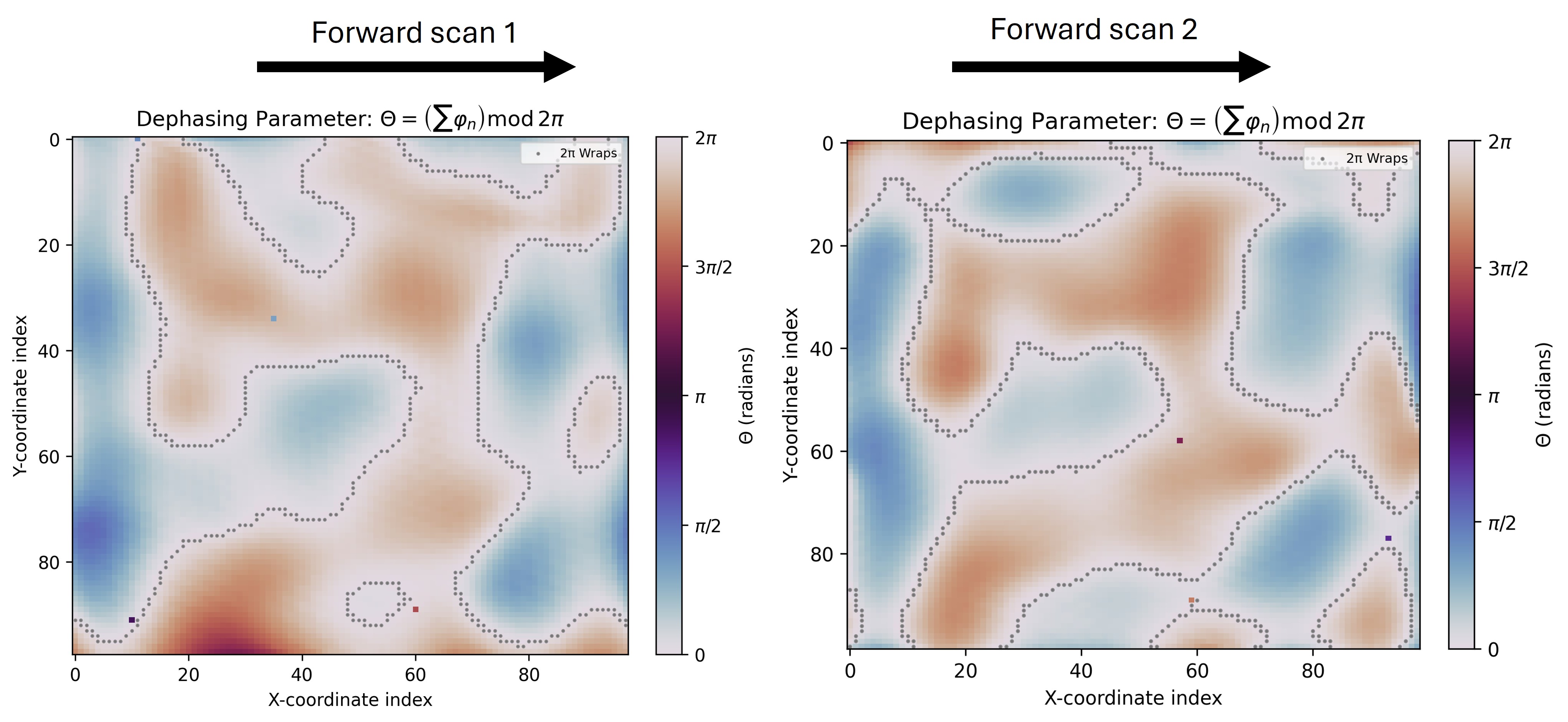}
    \caption{
        \textbf{Repeated STM scanning dephasing maps (forward fast axis $+x$, slow axis $+y$}. The repeated area scanning shows local changes in the dephasing parameter and structural evolution of the soliton (gray lines), indicating a glass like behavior.
    }
    \label{fig:tip_repeated_scanning}
\end{figure}

\newpage

\begin{figure}[h!]
    \centering
    \includegraphics[width=0.9\textwidth]{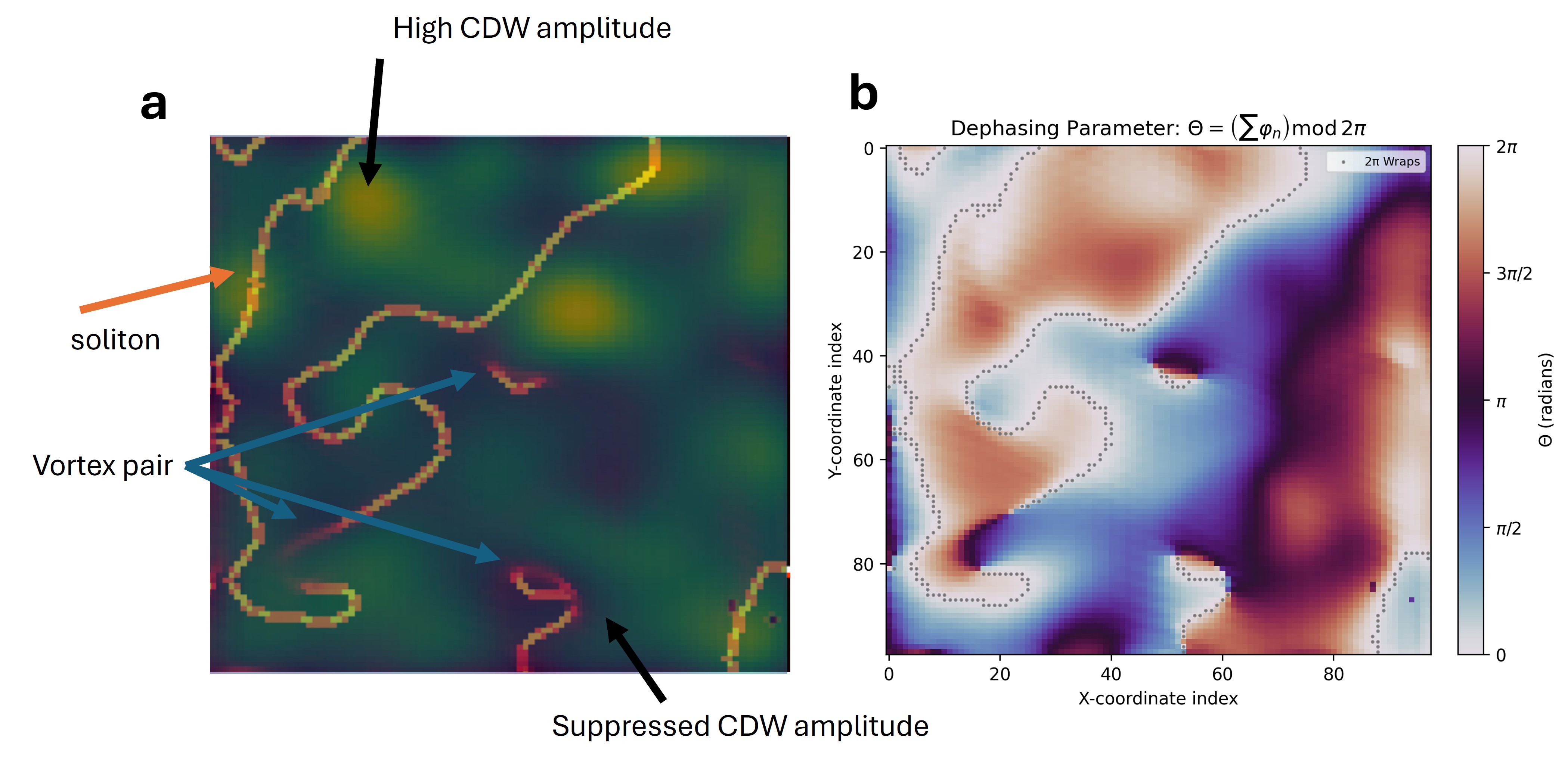}
    \caption{
        \textbf{Overlay of phase discommensuration or solitons (dark lines) on top of the CDW amplitude map sum ($A_1 + A_2 + A_3$) of the CDW order parameter.} (\textbf{b}) The accompanying dephasing map $\Theta$ is given. Soliton and vortices accompany suppressed CDW amplitude, especially the confined vortex-anti-vortex pairs.  
    }
    \label{fig:Supp_amplitude_solitons}
\end{figure}

\newpage
\section{Model of topological defect self-organization and pinning near structural defects}
\label{Supp_sec:fractality_model}

To model the self-organization of discommensuration or solitons, we conducted morphological analyses to establish a quantitative link between the observed soliton networks and the underlying influence of material defects and pinning potentials.

First, we investigated the scale-invariance of the soliton network. We observe that the solitons form a fibrous, percolating-like pathway, a morphology often associated with self-organization in strongly-correlated materials like {LaSrMnO$_3$} \cite{zotero-item-2692}. To quantify this, we analyzed the relationship between the perimeter $P$ and area $A$ for the individual phase domains enclosed by the solitons. For a fractal boundary, these quantities scale according to the relation $P \propto A^{D_b/2}$, where $D_b$ is the boundary fractal dimension.

\begin{figure}[h!]
    \centering
    \includegraphics[width=\textwidth]{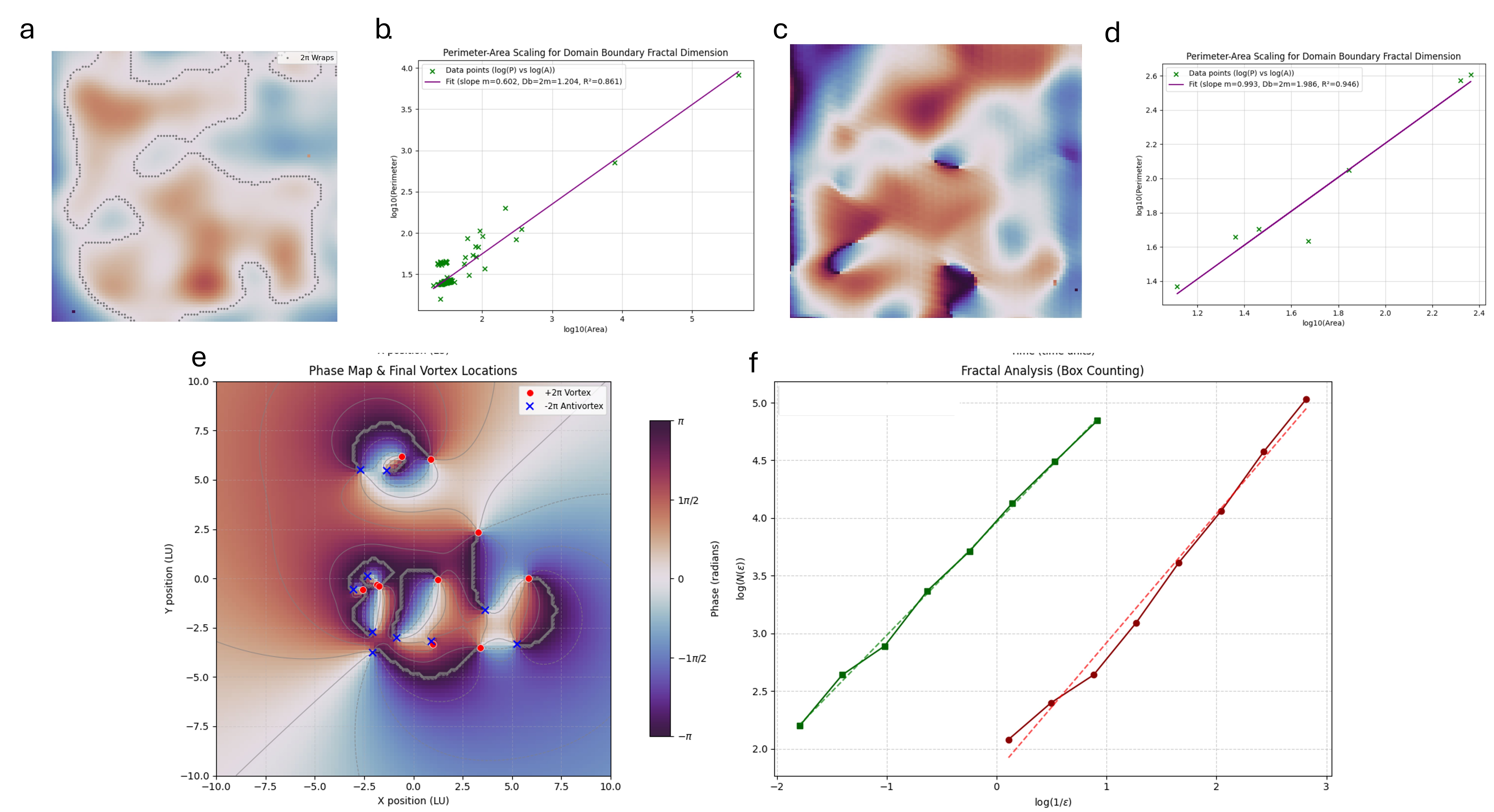} 
    \caption{\textbf{Experimental analysis of soliton boundary morphology.}
    (\textbf{a, b}) Dephasing map and perimeter–area (P-A) scaling for an area large soliton with no vortices. The fit yields a boundary fractal dimension of $D_b \approx 1.20$, indicating moderate fractality.
    (\textbf{c, d}) Dephasing map and P-A scaling for a soliton network near a single structural defect. The fit yields $D_b \approx 1.99$, demonstrating a transition to a highly complex, space-filling fractal boundary due to local disorder.
   (\textbf{e}) Simulated equilibrium of vortex pairs and connected solitons in the phase map. 
   (\textbf{f}) Extracted soliton (DW pairs)  (green) and phase contour (red) via the box-counting approach, showing fractal dimensions around 1.12.}
    \label{fig:supp_fractal}
\end{figure}

We performed this analysis on two distinct regions of the sample. In a region with low vortex density, but high structural defect density and a typical large soliton network, the P-A scaling yields a boundary dimension of \textbf{$D_b = 1.20 \pm 0.05$} (\textbf{Figure ~\ref{fig:supp_fractal}a}). This value, clearly greater than 1, indicates that the soliton boundaries are moderately fractal—more complex than a simple line, but not space-filling. This suggests a measurable self-affine roughness.

In stark contrast, the analysis of a soliton and vortex network in the immediate vicinity of a large single structural defect reveals a dramatically different morphology (\textbf{Figure ~\ref{fig:supp_fractal}b}). Here, the P-A scaling gives a boundary dimension of \textbf{$D_b = 1.99 \pm 0.06$}. A dimension of nearly 2 signifies a space-filling boundary of maximum tortuosity. This result implies that the local disorder from the defect forces the soliton network into a more complex, fractal state. 

\subsection*{Phenomenological Langevin model}

To test the pinning induced self-organisation of solitons, we developed a phenomenological model based on the Langevin equation \cite{Matsuura2015ChargeLiquid, Brazovskii1993CurrentWaves} to simulate the soliton dynamics. This overdamped approach we note is likely well-suited for CDW systems where viscous drag from coupling to the lattice is strong \cite{Singh2022Latticedriven2}. The soliton is discretized into a chain of $N$ beads connected by springs, representing a flexible domain wall. The equation of motion for the $i$-th bead at position $\vec{r}_i$ is given by:
$$
\eta \frac{d\vec{r}_i}{dt} = \vec{F}_{\text{spring},i} + \vec{F}_{\text{pin},i} + \vec{F}_{\text{drive},i} + \vec{F}_{\text{noise},i}(t)
$$
Here, $\eta$ is a viscous friction coefficient. The motion of each bead is determined by the sum of four forces:
\begin{itemize}
    \item \textbf{Internal elastic force ($\vec{F}_{\text{spring},i}$):} Maintains the soliton's integrity via Hookean springs (constant $k_{\text{DW}}$, rest length $l_0$) connecting adjacent beads. This term represents the soliton's line tension.

    \item \textbf{Pinning force ($\vec{F}_{\text{pin},i}$):} Describes the interaction with the disordered material landscape. It is derived from a static potential, $\vec{F}_{\text{pin},i} = -\nabla U(\vec{r}_i)$, constructed from a random spatial distribution of Gaussian potential wells representing defects.

    \item \textbf{External driving force ($\vec{F}_{\text{drive},i}$):} A uniform external force, such as from an electric field.

\end{itemize}

A simulation of solitons in a disordered potential (\textbf{Figure ~\ref{fig:supp_fractal}e-f}) yields a discommensuration with a fractal dimension of \textbf{$D \approx 1.12$}. This is in agreement with our experimental measurement of the typical soliton network, likely indicating that pinning influences the self-organization into a fractal structure of soliton networks.

\newpage
\section{Derivation of the phase chiral order parameter}
\label{Supp_sec:derivition_chiral_order}

For a 3Q CDW, the local charge modulation $\delta\rho(\mathbf{r})$ is the real part of the sum of three complex order parameters, $\Delta_j = A_j e^{i\phi_j}$ for $j=1, 2, 3$. We define a local scalar chiral order parameter, $\chi$, that quantifies the handedness of the phase relationships between these components. This parameter must satisfy several key symmetry requirements:

\begin{enumerate}
    \item \textbf{Gauge invariance:} $\chi$ must be invariant under a global phase shift, where each phase is transformed as $\phi_j \rightarrow \phi_j + \alpha$. 
    
    \item \textbf{Chiral nature:} $\chi$ must be odd under a spatial symmetry operation that inverts chirality. For a hexagonal lattice, a relevant operation is a mirror reflection, $\mathcal{M}$, which corresponds to an anti-cyclic permutation of the component indices (e.g., $1 \to 3 \to 2 \to 1$). Therefore, we require $\mathcal{M}(\chi) = -\chi$.
    
    \item \textbf{Scalar under lattice rotations:} As a scalar quantity, $\chi$ must be invariant under the discrete rotational symmetries of the hexagonal lattice. A key operation is a C$_3$ rotation, which corresponds to a cyclic permutation of the indices ($1 \to 2 \to 3 \to 1$). 
 \end{enumerate}

We construct the simplest expression that satisfies the above constraints; $\Delta_i^* \Delta_j = A_i A_j e^{i(\phi_j - \phi_i)}$. To satisfy the requirement of scalar behavior under lattice rotation (constraints), we must combine these terms in a way that is invariant under the cyclic permutation of indices. This leads to the following complex parameter as the simplest candidate:
$$ C = \Delta_1^* \Delta_2 + \Delta_2^* \Delta_3 + \Delta_3^* \Delta_1 $$.

We then enforce the chiral property (constraint 2) by examining how $C$ transforms under a mirror reflection $\mathcal{M}$, which acts as an anti-cyclic permutation ($1 \to 3$, $2 \to 2$, $3 \to 1$).
\begin{align*}
\mathcal{M}(C) &= \mathcal{M}(\Delta_1^* \Delta_2 + \Delta_2^* \Delta_3 + \Delta_3^* \Delta_1) \\
&= \Delta_3^* \Delta_2 + \Delta_2^* \Delta_1 + \Delta_1^* \Delta_3 \\
&= (\Delta_2^* \Delta_3)^* + (\Delta_1^* \Delta_2)^* + (\Delta_3^* \Delta_1)^* \\
&= (\Delta_1^* \Delta_2 + \Delta_2^* \Delta_3 + \Delta_3^* \Delta_1)^* = C^*
\end{align*}

Under the mirror operation, $C$ transforms into its complex conjugate. A complex quantity $C = \text{Re}(C) + i\,\text{Im}(C)$ that transforms this way has a real part that is even under the operation ($\mathcal{M}(\text{Re}(C)) = \text{Re}(C)$) and an imaginary part that is odd ($\mathcal{M}(\text{Im}(C)) = -\text{Im}(C)$). To isolate the purely chiral component, we define our order parameter $\chi$ as the imaginary part:

$$ \chi = \text{Im}(C) = \text{Im}(\Delta_1^* \Delta_2 + \Delta_2^* \Delta_3 + \Delta_3^* \Delta_1) $$

Substituting $\Delta_j = A_j e^{i\phi_j}$, we obtain the final expression in terms of experimentally measurable amplitudes and phases:

\begin{equation}
\chi(\mathbf{r}) = A_1A_2\sin(\phi_2-\phi_1) + A_2A_3\sin(\phi_3-\phi_2) + A_3A_1\sin(\phi_1-\phi_3)
\end{equation}

This expression is non-zero only if the phases have a net winding, making it a direct measure of local CDW SoD rotation handedness.


\newpage
\section{Symmetry analysis of the phase chiral order parameter}
\label{Supp_sec:symmetry_chiral_order}

\subsection*{\textbf{Point group operations}}

\paragraph{Rotations ($C_n$)}

The principal $z$-axis rotations act by permuting the in-plane wave vectors and their corresponding order parameters.

\begin{itemize}
    \item \textbf{$C_3$ (120$^\circ$) rotation:} The threefold rotation induces a cyclic permutation of the wave vectors $\Delta_1 \to \Delta_2 \to \Delta_3 \to \Delta_1$.
    \begin{align}
        C_3(\chi) &= A_2A_3\sin(\phi_3-\phi_2) + A_3A_1\sin(\phi_1-\phi_3) + A_1A_2\sin(\phi_2-\phi_1) \nonumber \\
        &= \chi
    \end{align}
    The parameter is even under $C_3$, consistent with the trigonal symmetry of the lattice.

    \item \textbf{$C_2$ (180$^\circ$) rotation:} This operation inverts the in-plane wave vectors, $\mathbf{q}_j \to -\mathbf{q}_j$, which maps the order parameters to their complex conjugates, $\Delta_j \to \Delta_j^*$. This corresponds to a phase inversion, $\phi_j \to -\phi_j$.
    \begin{align}
        C_2(\chi) &= A_1A_2\sin(-\phi_2 - (-\phi_1)) + A_2A_3\sin(-\phi_3 - (-\phi_2)) + A_3A_1\sin(-\phi_1 - (-\phi_3)) \nonumber \\
        &= -\left[ A_1A_2\sin(\phi_1-\phi_2) + A_2A_3\sin(\phi_2-\phi_3) + A_3A_1\sin(\phi_3-\phi_1) \right] \nonumber \\
        &= -\chi
    \end{align}
    The parameter is odd under $C_2$.
\end{itemize}

\paragraph{Mirror reflections ($\sigma$)}
\begin{itemize}
    \item \textbf{Vertical mirror plane ($\sigma_v$):} A mirror plane aligned with a wave vector (e.g., $\mathbf{q}_1$) leaves that component invariant while swapping the other two: $\Delta_1 \to \Delta_1$, $\Delta_2 \leftrightarrow \Delta_3$.
    \begin{align}
        \sigma_v(\chi) &= A_1A_3\sin(\phi_3-\phi_1) + A_3A_2\sin(\phi_2-\phi_3) + A_2A_1\sin(\phi_1-\phi_2) \nonumber \\
        &= -\left[ A_3A_1\sin(\phi_1-\phi_3) + A_2A_3\sin(\phi_3-\phi_2) + A_1A_2\sin(\phi_2-\phi_1) \right] \nonumber \\
        &= -\chi
    \end{align}
    The parameter is odd under $\sigma_v$, confirming its chiral nature (breaking of mirror symmetry).

    \item \textbf{Horizontal mirror plane ($\sigma_h$):} This operation reflects the coordinate $z \to -z$. Since the CDW is defined by in-plane wave vectors, the order parameters are invariant under this operation.
    \begin{equation}
        \sigma_h(\chi) = \chi
    \end{equation}
    The parameter is even under $\sigma_h$.
\end{itemize}

\paragraph{Inversion ($i$)}
Spatial inversion ($\mathbf{r} \to -\mathbf{r}$) is equivalent to a $C_2$ rotation followed by a horizontal reflection ($i = \sigma_h \circ C_2$).
\begin{equation}
    i(\chi) = \sigma_h(C_2(\chi)) = \sigma_h(-\chi) = -\chi
\end{equation}
The parameter is odd under inversion (P-odd). This implies that $\chi$ describes a true chiral state (e.g., a helix) rather than a pure (bulk) axial rotation.

\subsection*{Time reversal ($\mathcal{T}$)}

The time reversal operator $\mathcal{T}$ acts by complex conjugation on the order parameters, $\Delta_j \to \Delta_j^*$, which is equivalent to inverting the phases, $\phi_j \to -\phi_j$. The transformation is therefore identical to that of the $C_2$ rotation.

\begin{equation}
    \mathcal{T}(\chi) = -\chi
\end{equation}

The parameter is odd under time reversal (T-odd).

\begin{table}[h]
\centering
\caption{Summary of transformations for the chiral order parameter $\chi$.}
\begin{tabular}{l c c}
\hline\hline
Symmetry Operation & Representative & Transformation \\ [0.5ex] 
\hline
Six-fold rotation & $C_6$ & $\chi \to +\chi$ \\
Two-fold rotation & $C_2$ & $\chi \to -\chi$ \\
Vertical reflection & $\sigma_v$ & $\chi \to -\chi$ \\
Horizontal reflection & $\sigma_h$ & $\chi \to +\chi$ \\
Spatial inversion & $i$ & $\chi \to -\chi$ \\
Time reversal & $\mathcal{T}$ & $\chi \to -\chi$ \\ [1ex] 
\hline
\end{tabular}
\label{tab:symmetry}
\end{table}

While $\chi$ is odd under the time-reversal operation, the physical quantity measured in STM is the local charge density $\rho(\mathbf{r})$. This is a (within approximation) static observable and is therefore strictly T-odd. The sign of $\chi$ determines the sense of rotation (clockwise vs. counter-clockwise) of the CDW superlattice relative to the atomic lattice. Thus, while STM cannot directly detect time-reversal symmetry breaking (e.g., loop currents) without spin-polarization, it directly images the spatial enantiomers defined by the sign of $\chi$.

We must reconcile the symmetry constraints of the bulk crystal with the specific environment of our surface measurements. In the bulk, 1T-TaS$_2$ is characterized by ferro-rotational order ($\Omega$), an axial state which preserves inversion symmetry but breaks mirror symmetry \cite{Luo2021UltrafastGeneration, Liu2023ElectricalCrystals}. Consequently, the bulk state is non-polar and strictly forbids linear coupling to electric fields. In contrast, the chiral order parameter defined in our work, $\chi$, represents a distinct phase winding that is inversion-odd. While such a state is symmetry-forbidden in the centrosymmetric bulk, the crystal surface inherently breaks inversion symmetry. This symmetry breaking lifts the distinction between rotational and chiral representations, allowing the ferro-rotational order to induce a surface-localized chiral winding $\chi$. This distinction dictates how electric fields couple to the order parameter. An out-of-plane field ($E_z$) can couple to the surface state via an improper ferroelectric mechanism ($F \propto E_z \chi \Omega$), allowing for thermodynamic selection of the chiral domain. However, for an in-plane electric field ($E_{\parallel}$), the remaining $C_3$ rotational symmetry of the lattice forbids any static linear coupling ($P_{\parallel}=0$). Therefore, a static in-plane field cannot switch the global chirality through purely thermodynamic energy minimization.


\newpage
\section{The amplitude-weighted phase triangle model}
\label{Supp_sec:phase_triangle_chiral_order}

To visualize and quantify chirality in the three-component CDW, we developed an amplitude-weighted phase triangle model. Each of the three complex order parameters, $\Delta_i = A_i e^{i\phi_i}$, can be represented as a vector in the complex plane with its endpoint at the Cartesian coordinates $P'_i = (x'_i, y'_i) = (A_i\cos\phi_i, A_i\sin\phi_i)$. The triangle formed by the three vertices $P'_1, P'_2, P'_3$ provides a complete geometric representation of the local chiral state. 

We demonstrate here that the signed area of this triangle (Area') is directly proportional to the scalar chirality parameter, $\chi = \text{Im}(\Delta_1^* \Delta_2 + \Delta_2^* \Delta_3 + \Delta_3^* \Delta_1)$. We calculate the area using the shoelace formula for a polygon:
\begin{align*}
    \text{Area'} &= \frac{1}{2} \left[ x'_1(y'_2 - y'_3) + x'_2(y'_3 - y'_1) + x'_3(y'_1 - y'_2) \right] \\
    \text{Substituting } x'_i &= A_i\cos\phi_i \text{ and } y'_i = A_i\sin\phi_i: \\
    2 \times \text{Area'} &= A_1\cos\phi_1(A_2\sin\phi_2 - A_3\sin\phi_3) + A_2\cos\phi_2(A_3\sin\phi_3 - A_1\sin\phi_1) \\
    & \quad + A_3\cos\phi_3(A_1\sin\phi_1 - A_2\sin\phi_2) \\
    \text{Grouping terms }&\text{by amplitude pairs:} \\
    &= A_1A_2(\sin\phi_2\cos\phi_1 - \cos\phi_2\sin\phi_1) + A_2A_3(\sin\phi_3\cos\phi_2 - \cos\phi_3\sin\phi_2) \\
    & \quad + A_3A_1(\sin\phi_1\cos\phi_3 - \cos\phi_1\sin\phi_3) \\
    \text{Using the identity }& \sin(\alpha - \beta) = \sin\alpha\cos\beta - \cos\alpha\sin\beta: \\
    &= A_1A_2\sin(\phi_2 - \phi_1) + A_2A_3\sin(\phi_3 - \phi_2) + A_3A_1\sin(\phi_1 - \phi_3)
\end{align*}

This final expression is identical to our definition of the scalar chirality, $\chi$. Therefore, we establish the direct relationship:
$$ \chi = 2 \times \text{Area}' $$

The sign of the triangle's area directly corresponds to the system's handedness. A positive area ($\chi > 0$) results from a counter-clockwise ordering of the vertices ($P'_1 \to P'_2 \to P'_3$), while a negative area ($\chi < 0$) results from a clockwise ordering. An achiral state ($\chi = 0$) occurs when the three vertices become collinear and the triangle's area vanishes. This transition is expected at domain walls that separate regions of opposite chirality.

\newpage
\section{STM simulation of the NC-CDW and chirality}
\label{Supp_sec:chirality_simulation_NCCDW}

\begin{figure}[h!]
    \centering
    \includegraphics[width=0.9\textwidth]{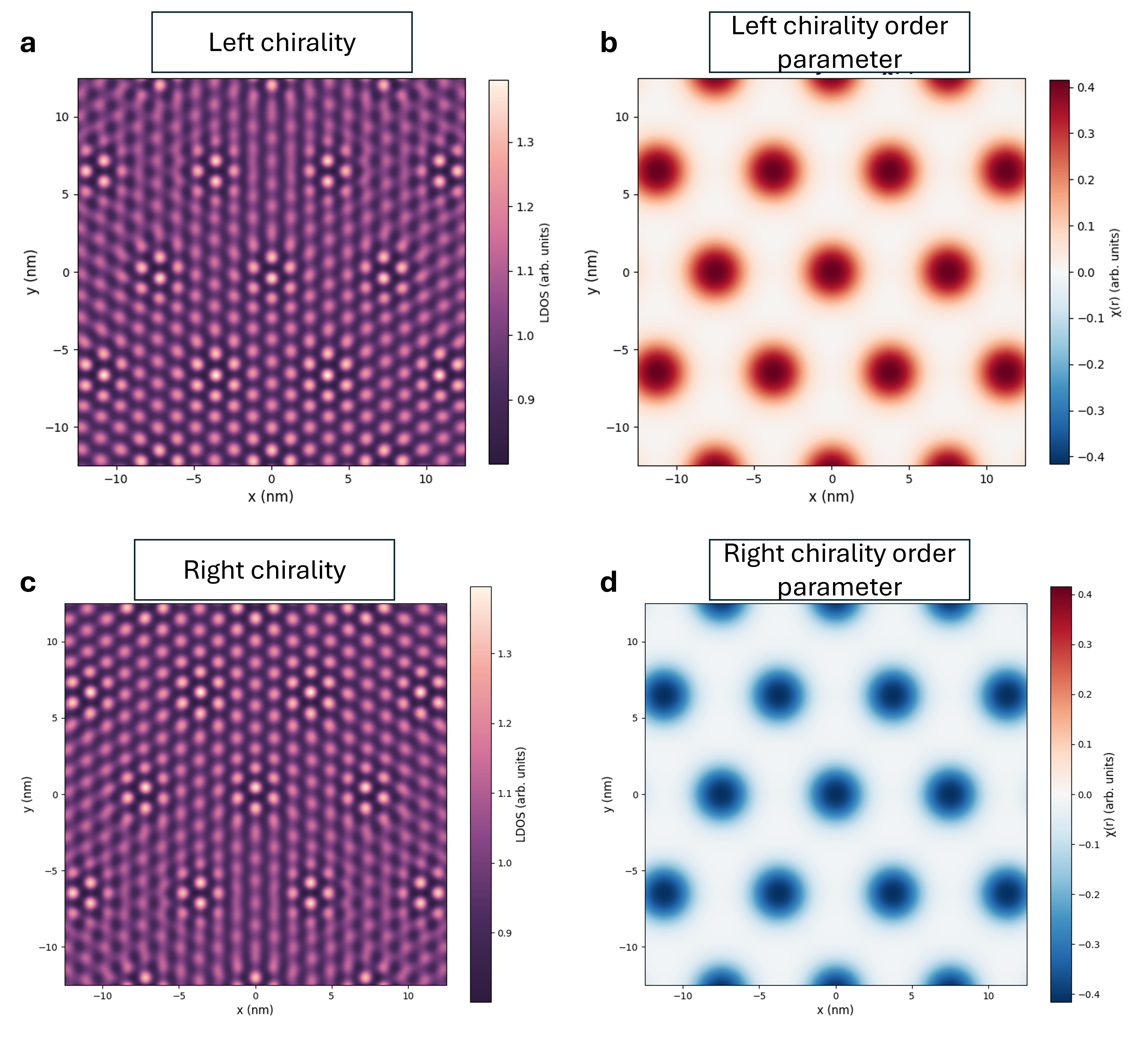}
    \caption{
        \textbf{STM simulation of two chiral NC-CDW phases (left and right) and the accompanying chiral order map $\chi$ based on the relative phases}. Note the clear chiral inversion of the NC-CDW domains \cite{Singh2022Latticedriven2} and reduced winding strength in the domain walls. 
    }
    \label{fig:Simulation_chiral}
\end{figure}

\newpage
\section{Topological classification of decoupled chiral and topological phase orders}
\label{Supp_sec:topology_classification}

We describe the topological defect landscape of the CDW by two independent, coexisting order parameter fields: a chiral field and continuous CDW order phase fields. The stability and classification of the distinct defects associated with each field are determined by separate homotopy groups \cite{Skogvoll2023UnifiedExcitations}, which explains their observed physical independence and the spatial decoupling of their associated defect structures.

\subsection*{The chiral order and zeroth homotopy group}
The chiral order parameter is constructed from the complex CDW amplitudes $\Delta_i$ ($i=1,2,3$). This quantity is invariant under global phase shifts but changes sign under permutations of the wavevectors, making it sensitive to the handedness of the CDW structure. The system spontaneously breaks discrete symmetry by selecting one of two chiral ground states: positive ($\chi > 0$) or negative ($\chi < 0$), where $\chi = \text{Im}(C)$ serves as the scalar chiral order parameter. The order parameter space (OPS) for this discrete choice is a two-point set, $V_{chiral} = \{v_+, v_-\}$. 

Defects that form boundaries between these distinct chiral states are classified by the zeroth homotopy group, $\pi_0(V)$, which enumerates the path-connected components of the space:
$$
\pi_0(V_{chiral}) \cong \mathbb{Z}_2
$$
A non-trivial zeroth homotopy group necessitates the existence of topologically stable domain walls separating regions of opposite chirality. Experimentally, these correspond to line defects where the scalar chiral order parameter vanishes, $\chi(\mathbf{r}) \approx 0$, marking the transition between $\chi > 0$ and $\chi < 0$ domains.

\subsection*{The phase order and its distinct defects}
Coexisting with, but independent from, the chiral order are the continuous phase fields $\phi_i(\mathbf{r})$ ($i=1,2,3$) associated with each CDW component, where $\Delta_i(\mathbf{r}) = |\Delta_i(\mathbf{r})| e^{i\phi_i(\mathbf{r})}$. Each phase field has order parameter space $V_{phase} = S^1$. Critically, the chiral order parameter $C$ depends on the \textit{relative} phases between components (through products like $\Delta_1^* \Delta_2 = |\Delta_1||\Delta_2| e^{i(\phi_2 - \phi_1)}$), while defects in the phase fields can involve changes in individual phases or their sum. This mathematical structure allows phase defects to exist independently of chiral domain walls as they are governed by distinct topological classifications:

\subsubsection*{1. Point defects (vortices)}
Point defects in a 2D phase field are classified by the first homotopy group:
$$
\pi_1(V_{phase}) = \pi_1(S^1) \cong \mathbb{Z}
$$
This guarantees the existence of stable 0D defects (vortices and anti-vortices) in the individual phase fields $\phi_i$, characterized by an integer winding number, $n \in \mathbb{Z}$. These defects are topologically protected and can serve as endpoints when phase slip lines (branch cuts) connect vortex-anti-vortex pairs. Because the chiral order depends on \textit{relative} phases $(\phi_j - \phi_i)$ rather than absolute phases, a vortex in a single component (where $\phi_i$ winds by $2\pi$) does not necessarily create a chiral domain wall.

\subsubsection*{2. Line defects (phase discommensurations and solitons)}
Line defects (phase slip lines) in a single phase field are classified by the zeroth homotopy group:
$$
\pi_0(V_{phase}) = \pi_0(S^1) \cong \mathbb{Z}_1 \text{ (the trivial group)}
$$
Because this group is trivial (has only one component), a closed $2\pi$ soliton loop in a single phase component is in principle not topologically protected. 

However, the three-component CDW system exhibits multiple classes of line structures:

\textbf{(i) NC discommensuration network:} In the NC-CDW phase, the system forms a network of domain walls where all three components simultaneously undergo fractional $2\pi/3$ phase slips: $\phi_i \to \phi_i + 2\pi/3$ for all $i$. While each individual slip is not topologically protected by homotopy theory. The $2\pi$ phase wraps in $\Theta = \sum_i \phi_i$ trace this intrinsic NC matrix. Crucially, these discommensurations preserve the relative phases between components. The NC matrix can exist throughout regions of uniform chirality.

\textbf{(ii) $\pi$ solitons:} these solitons connect vortex-anti-vortex pairs via the shortest pathway to reduce line tension and minimize energy via CDW amplitude supression. 

\subsection*{Topological decoupling and spatial independence}
The distinct homotopy groups governing chiral domain walls ($\pi_0(\mathbb{Z}_2)$) and phase defects ($\pi_1(S^1)$ for vortices, energetically stabilized for line defects) can show that these defect types are topologically decoupled. This decoupling has experimental consequences:

\begin{itemize}
    \item \textbf{Spatial separation:} Chiral domain walls (where $\chi = 0$) and phase discommensurations (where $\Theta$ exhibits $2\pi$ wraps) need not coincide spatially. We observe extensive regions where the NC discommensuration network exists within uniform chiral domains, confirming their independence.
    
    \item \textbf{Independent dynamics:} The two defect types can nucleate, move, and annihilate independently. A phase discommensuration can propagate through regions of uniform chirality without inducing chiral reversal, and vice versa; a chiral domain wall can move without necessarily creating phase defects.
\end{itemize}

\subsection*{Decoupling and coupling of phase and chiral defects: Two regimes}

The spatial relationship between phase discommensurations and chiral domain walls depends critically on the symmetry of the phase shifts at the domain wall. We distinguish two fundamentally different regimes:

\textbf{(i) Symmetric discommensurations (decoupled):} Consider a configuration where all three phase components shift by equal amounts across a domain wall. In Region A ($x < 0$), the phases are $(\phi_1, \phi_2, \phi_3) = (0, 0, 0)$, while in Region B ($x > 0$), they are $(\phi_1, \phi_2, \phi_3) = (2\pi/3, 2\pi/3, 2\pi/3)$. The dephasing map exhibits a $2\pi$ phase wrap: $\Theta_A = 0$ and $\Theta_B = (2\pi/3 + 2\pi/3 + 2\pi/3) \bmod 2\pi = 0$. However, the chiral order remains uniform. For uniform amplitudes $|\Delta_i| = \Delta_0$, we find $C_A = \Delta_0^2 [e^{i(0)} + e^{i(0)} + e^{i(0)}] = 3\Delta_0^2$ and $C_B = \Delta_0^2 [e^{i(0)} + e^{i(0)} + e^{i(0)}] = 3\Delta_0^2$ (since all relative phases $\phi_j - \phi_i$ vanish in both regions), giving $\chi_A = \chi_B$. This represents the equilibrium NC discommensuration network, where the domain walls preserve three-fold rotational symmetry. Such symmetric discommensurations can propagate through regions of uniform chirality without inducing chiral inversion, demonstrating complete topological decoupling.

\textbf{(ii) Asymmetric discommensurations (coupled):} In contrast, consider a partial discommensuration where only one or two phase components shift. For example, Region A has $(\phi_1, \phi_2, \phi_3) = (0, 2\pi/3, 4\pi/3)$ (right-handed star configuration), while Region B has $(\phi_1, \phi_2, \phi_3) = (2\pi/3, 4\pi/3, 4\pi/3)$ (with only $\phi_1$ and $\phi_2$ shifted relative to a left-handed configuration). The chiral order now differs across the boundary. In Region A, the relative phases are $(\phi_2 - \phi_1, \phi_3 - \phi_2, \phi_1 - \phi_3) = (2\pi/3, 2\pi/3, 2\pi/3)$, yielding positive chirality $\chi_A > 0$. In Region B, after calculating the relative phases and evaluating $C_B = \Delta_0^2 [e^{i(\phi_2-\phi_1)} + e^{i(\phi_3-\phi_2)} + e^{i(\phi_1-\phi_3)}]$ with the asymmetric phase values, one obtains different or even opposite chirality $\chi_B \neq \chi_A$ (and potentially $\chi_B < 0$). Such asymmetric discommensurations, likely induced by disorder or external fields, break the three-fold symmetry differently on each side of the domain wall, thereby changing the relative phases that determine chirality. These defects represent bound states where the phase discommensuration is intrinsically coupled to a chiral domain wall, and their motion necessarily transports chiral inversion fronts.

\newpage

\begin{figure}[h]
    \centering
    \includegraphics[width=\textwidth]{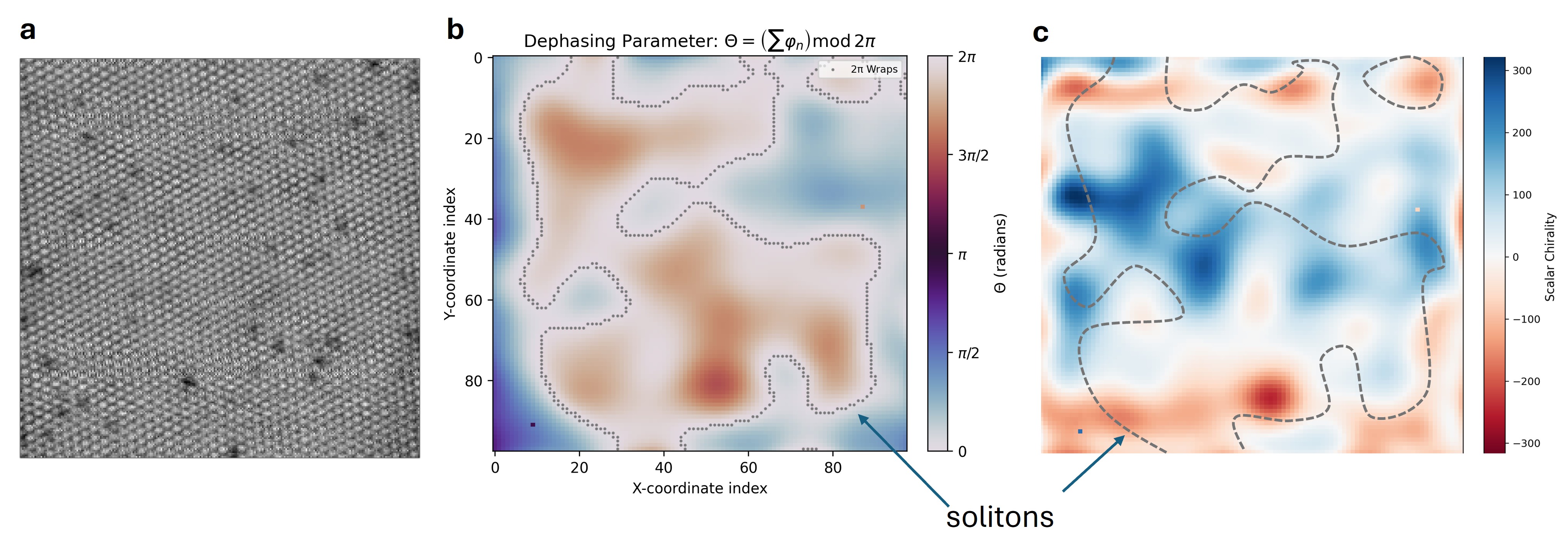}
    \caption{\textbf{Comparison between STM map, soliton map and chiral order map.} (\textbf{a}) STM height mapping showing a structural defect rich area. Same map as Figure \ref{fig:chirality_model}a of the main text. (\textbf{b}) Dephasing map $\Theta$ where the gray lines indicate meandering solitons with $2\pi$ phase slips. (\textbf{c}) Accompanying chiral order, overlaid with the phase slip network; they do not spatially coincide.}
    \label{topology_decoupling}
\end{figure}

\newpage
\subsection*{S10: Chiral inversion model}
\label{Supp_sec:chiral_inversion_model}

We defined a phenemenological model based on a simple Ginzburg-Landau model, where we have a $\phi^4$ double well potential and coupling to the electric field driving discommensurations or soliton movement. By no means does it describe all the CDW GL components \cite{Xu2020SpontaneousDichalcogenide}, but provides a tentative insight into symmetry breaking under soliton movement. Future work can extend the rigoursness of this model. In this model, the soliton melts the local CDW amplitude. The three CDW phases $\phi_1, \phi_2, \phi_3$ are reconstructed from a collective coordinate $\Sigma$ and relative coordinates $\Delta_{ij}$:
\begin{align}
    \Sigma(\mathbf{r}, t) &= \sum_{i=1}^3 \phi_i \quad \text{(Phase field)} \\
    \Delta_{ij}(\mathbf{r}, t) &= \phi_j - \phi_i \quad \text{(Relative phase)}
\end{align}

The phase field $\Sigma$ governs the sliding motion and the discommensuration structure (soliton lattice). It evolves according to:
\begin{equation}
    \eta_\Sigma \frac{\partial \Sigma}{\partial t} = \underbrace{K_\Sigma \nabla^2 \Sigma}_{\text{Elasticity}} - \underbrace{V_{\text{pin}} \Psi^2 \sin(\Sigma)}_{\text{Pinning}} + \underbrace{\gamma E(t) \rho_{\text{eff}}}_{\text{Elec. Drive}}
\end{equation}
where $\rho_{\text{eff}} \propto \nabla \Sigma$ is the effective charge density. The CDW amplitude $\Psi$ is adiabatically ecoupled to the phase field gradient, creating "soft" cores at soliton locations:
\begin{equation}
    \Psi(\mathbf{r}) \approx \Psi_0 \left( 1 - \alpha |\nabla \Sigma|^2 \right)
\end{equation}

The chiral order parameter $\chi$ evolves in a double-well potential. The critical feature of this model is represented by a spatially variable mobility $\Gamma(\mathbf{r},t)$:
\begin{equation}
    \frac{\partial q}{\partial t} = \Gamma_{\text{boost}}(\mathbf{r},t) \cdot \frac{1}{\eta_q} \left[ \underbrace{(B_q q - A_q q^3)}_{\text{Landau Potential}} + K_q \nabla^2 q + \underbrace{D E(t)}_{\text{Bias}} \right]
\end{equation}
The boost factor $\Gamma_{\text{boost}}$ is defined such that chirality dynamics are frozen in the bulk (ferro rotational order does not allow for linear coupling of an in-plane E-field to the chiral order \cite{Liu2023ElectricalCrystals}, but accelerated (catalyzed) near moving solitons:
\begin{equation}
    \Gamma_{\text{boost}} \propto \left(1 - \frac{\Psi}{\Psi_0}\right) \left| \frac{\partial \Sigma}{\partial t} \right|
\end{equation}

The relative phases $\Delta_{ij}$ are elastically coupled to the chirality field $q$, enforcing the geometric Star-of-David pattern:
\begin{equation}
    \frac{\partial \Delta_{ij}}{\partial t} = - \frac{J_{\text{lock}}}{\eta_\Delta} \Psi^2 \left( \Delta_{ij} - \theta_{\text{target}}(q) \right)
\end{equation}

\subsubsection*{Numerical implementation notes}

Simulations were performed on a $200 \times 200$ spatial grid with periodic boundary conditions. Time integration used forward Euler with timestep $\Delta t = 0.04$ (dimensionless units). Laplacians were computed using five-point finite difference stencils. Phase gradients were computed with centered differences, properly handling $2\pi$ branch cuts. Gaussian filtering for the boost function used FFT-based convolution with kernel width $\sigma = 8$ grid points.

Initial conditions: Random chirality field smoothed to remove short-wavelength noise; random phase field seeded with $N=20$ vortex-anti-vortex pairs connected to phase jumps of $2\pi$ in the phase field.

\newpage
\section*{Materials and Methods}

\subsection*{Scanning tunneling microscopy}

Scanning tunneling microscopy was performed using a Scienta Omicron variable temperature SPM. The sample was grounded and the tip biased. Measurements were taken at 300K. We used a mechanically cut PtIr tip, which was calibrated on HOPG to achieve atomic resolution and LDOS calibration. Drift calibration was implemented by a custom algorithm; we track atomic positions on HOPG between consecutive scans. The carbon atoms are mapped by first applying a Gaussian filter (sigma 2.0) to the image and using the peak finding algorithm, implemented in Python. Communication between Python and the STM is enabled through the nOmicron module \cite{OGordon1002019OGordon100Conditioning}. Over several hours of scanning, the drift in the four scan directions (forward/backward and up/down) are mapped and stored after every second image. These transient drift directions are then used to calibrate the temporal drift and enable real-time drift correction. TaS$_2$ single crystals are obtained from 2D Semiconductors grown with flux method and cleaved in UHV at 300K using the scotch-tape method. Raw images are processed in Gwyddion \cite{Necas2012GwyddionAnalysis} by applying polynomial background subtraction and line alignment. Line alignment was used to identify maps that were likely influenced by tip changes, which were discarded for further analysis.

\subsubsection*{Scanner nonlinearities}

The exact movements of the STM tip are regulated by a scanner tube, which can display inherent nonlinear responses to voltage, particularly hysteresis and creep \cite{Yothers2017RealspaceImages}, which predominately cause distortions in images obtained during both forward (e.g., left-to-right) and backward (right-to-left) scanning paths [..]. We attempted to use the forward and backward scan to map changes in the dephasing parameter and chiral order, for example influence by the electric field and/or tunnel current \cite{Liu2023ElectricalCrystals, Song2022AtomicscaleSwitching}. To correct for these physical non-idealities, we tested a post-processing technique that treats the distortion as a geometric transformation (affine global distortion) between the forward and backward scan images. We first assume that the CDW is static, at least within the time frame of the STM map, i.e. around 10 min, so any differences between the two images are artifacts of the scanner's motion. The correction procedure begins by identifying reference points, or fiducial markers, present in both images. For atomically resolved STM images, the atoms themselves are ideal markers. The code identifies these atoms by finding local intensity maxima after applying a Gaussian filter to reduce noise. This yields two sets of coordinates: a reference set $\{ \mathbf{p}_i \} = \{ (x_i, y_i) \}$ from the forward scan, and a distorted set $\{ \mathbf{q}_i \} = \{ (u_i, v_i) \}$ from the backward scan.

\begin{figure}[h]
    \centering
    \includegraphics[width=\textwidth]{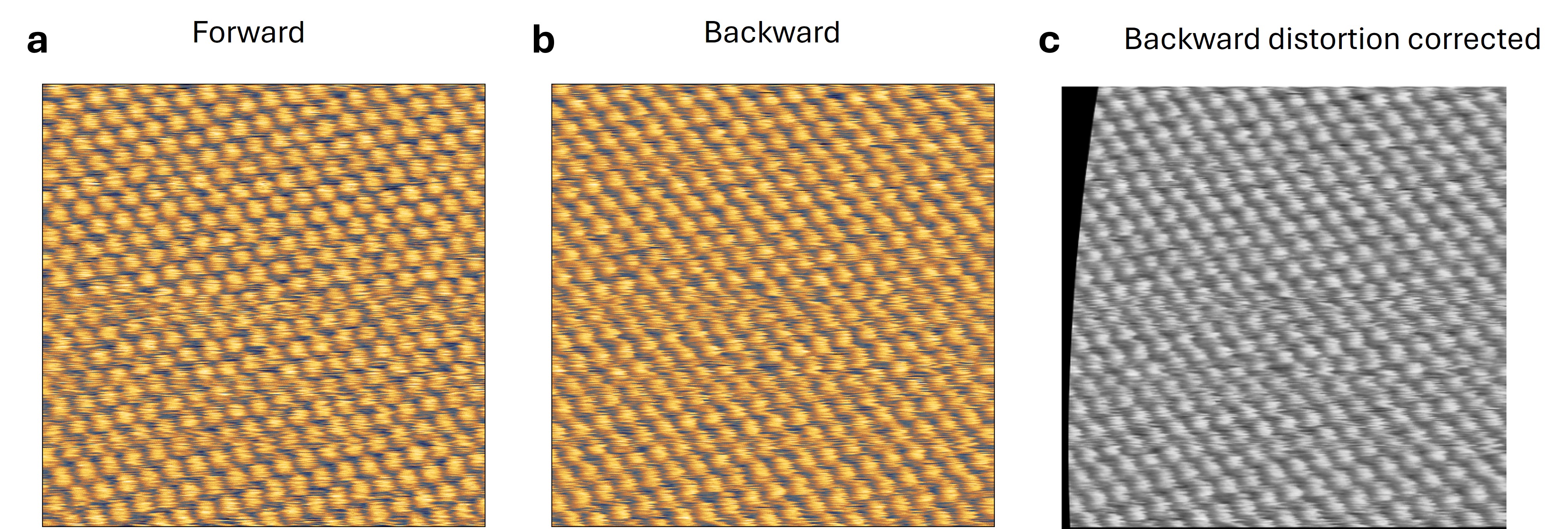}
    \caption{Forward $+x$ and backward scan $-x$ (both slow axis $+y$ measured on HOPG at 300K. The backward scan is affine global distortion corrected by mapping it onto the forward scan map. }
    \label{fig:supp_distortion_correction}
\end{figure}

Next, a one-to-one correspondence between the peaks is established, by constructing a k-d tree from the forward image peaks, allowing a fast nearest-neighbor search to find the most likely matching peak for each peak in the backward image. This results in a set of corresponding point pairs $(\mathbf{p}_i, \mathbf{q}_i)$. With these pairs, the core of the correction is to find a mathematical transformation $T$ that maps the distorted coordinate system to the reference system such that $\mathbf{p} = T(\mathbf{q})$. Since piezoelectric distortions are smooth, they can be well-approximated by a low-order polynomial function, which we model the mapping as:
$$
\begin{aligned}
x &= f(u, v) = \sum_{i=0}^{N} \sum_{j=0}^{N-i} a_{ij} u^i v^j \\
y &= g(u, v) = \sum_{i=0}^{N} \sum_{j=0}^{N-i} b_{ij} u^i v^j
\end{aligned}
$$

The unknown coefficients $\{ a_{ij}, b_{ij} \}$ are determined by a least-squares fit that minimizes the error between the actual reference peak positions $\mathbf{p}_i$ and the transformed positions $T(\mathbf{q}_i)$. Once the optimal transformation $T$ is found, it defines the complete distortion field. To create the final corrected image, the backward scan image is "warped" using this transformation. This involves calculating the intensity value for each pixel in a new blank image by finding its corresponding coordinate in the original backward image via the transformation $T$ and interpolating the surrounding pixel values. The result is a new backward-scan image where all features are geometrically aligned with the forward-scan image, \textbf{Figure \ref{fig:supp_distortion_correction}} taken on HOPG, effectively removing the scanner-induced distortions.

\begin{figure}[h]
    \centering
    \includegraphics[width=\textwidth]{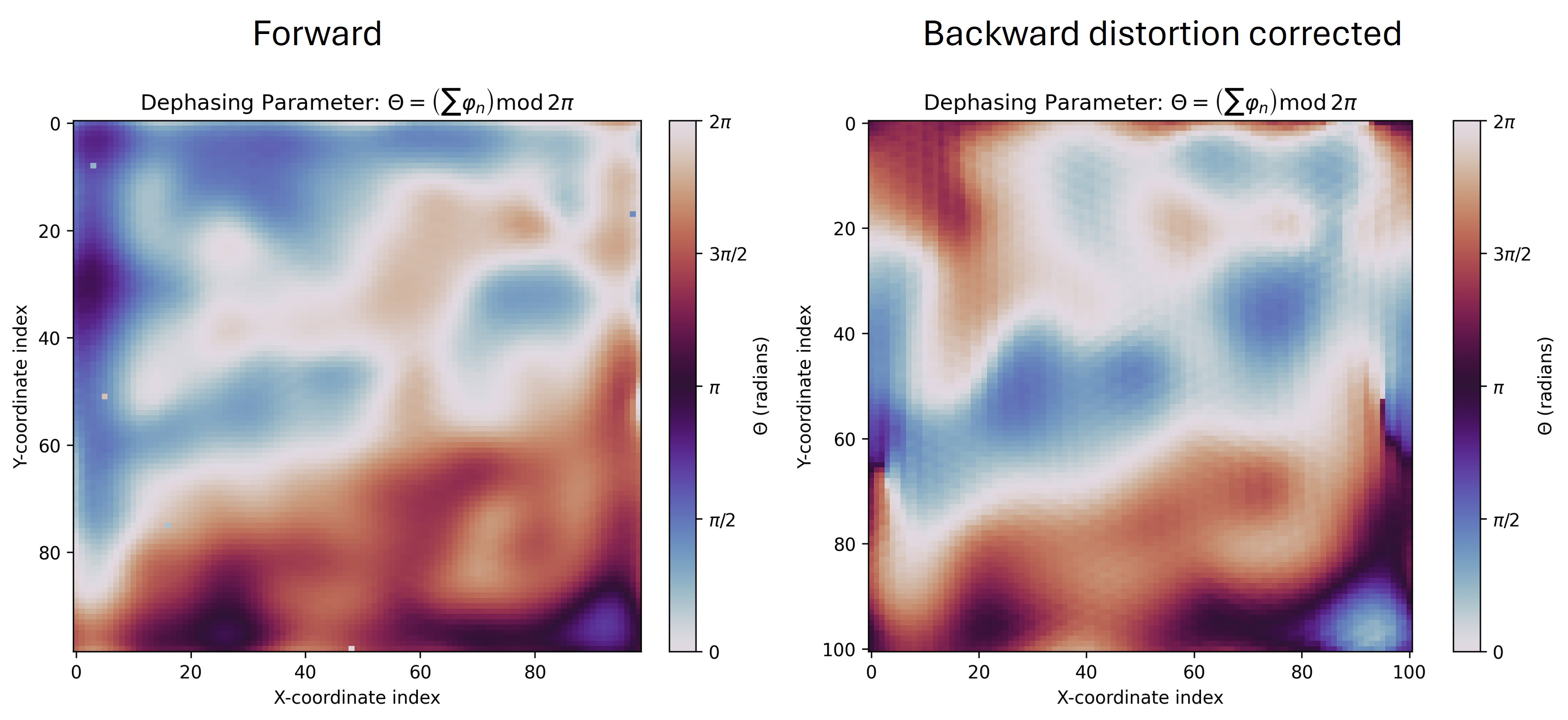}
    \caption{\textbf{Forward $+x$ and backward scan global affine corrected dephasing maps $-x$ (both slow axis $+y$ measured on TaS$_2$ at 300K.} The backward scan is affine global distortion corrected and shows only local variations in dephasing, which can be intrinsic perturbations of the CDW phase by the STM \cite{Song2022AtomicscaleSwitching} and not due to scanner non linearities.}
    \label{fig:supp_distortion_correction_TaS2}
\end{figure}

This distortion correction works fine for a simple lattice such as HOPG, the complex structure of the CDW of TaS$_2$ makes it more difficult. A globally distortion corrected and extracted dephasing maps is given in \textbf{Figure \ref{fig:supp_distortion_correction_TaS2}}, which shows local variations in the dephasing and position of solitons. However, we can still not be certain that these are still extremely small imperfections in the distortion corrections itself. Therefore, we only report STM maps (and subsequent data analysis) based on forward scans ($+x, +y$), similar to the work in Ref. \cite{Song2022AtomicscaleSwitching}. This means that there is still inherent scanner non linearity, but is normalized by the same scanner direction non-nonlinearities. We did also use this non-linear scanner analysis to note if datasets suffer from significant distortion, i.e. by thermal drift, and removed those images from the dataset.

\subsection*{Topological defect identification in dephasing maps}

The dephasing map, $\Theta(\mathbf{r})$, is analyzed to identify two distinct types of spatial features: phase slips and phase vortices. These features are located using numerical methods.

\subsubsection*{Phase slips or solitons}
Phase slips are detected by calculating the phase difference between adjacent grid points in both the horizontal and vertical directions. A phase slip is flagged at a grid point $(i, j)$ if the absolute value of the phase difference with any of its immediate neighbors, for instance $|\Theta_{i, j} - \Theta_{i-1, j}|$, approximates $2\pi$ within a defined tolerance (0.1 rad). This method effectively locates the boundaries where the phase value jumps from near $2\pi$ to near $0$, or vice versa.

\subsubsection*{Phase vortices}

Phase vortices are topological point defects in the dephasing map, characterized by a phase singularity where the value is undefined. The defining physical characteristic of a vortex is that the line integral of the phase gradient around any closed path $C$ enclosing the vortex core is quantized in integer multiples of $2\pi$:
$$
\oint_C \nabla \Theta \cdot d\mathbf{l} = 2\pi n
$$
where $n \in \mathbb{Z}$ is the topological charge of the vortex.

To identify these vortices numerically, the continuous line integral is discretized and computed over the smallest possible closed loop on the grid: a $2 \times 2$ pixel plaquette. For each plaquette in the dephasing map, the algorithm sums the wrapped phase differences between adjacent corner points in a counter-clockwise path:
$$
\Delta \Theta_{\text{loop}} = \Delta \Theta_{1 \to 2} + \Delta \Theta_{2 \to 3} + \Delta \Theta_{3 \to 4} + \Delta \Theta_{4 \to 1}
$$
The use of the wrapped phase difference, which maps the result to the interval $(-\pi, \pi]$, is crucial. It ensures that any $2\pi$ discontinuities arising from phase slips along the integration path do not contribute to the final sum, thereby isolating the contribution from the topological defect alone.

The topological charge $n$ for the plaquette is then calculated as $n = \Delta \Theta_{\text{loop}} / 2\pi$. A vortex with charge $n=+1$ or an anti-vortex with charge $n=-1$ is identified if the calculated charge is within a numerical tolerance of these integer values. The spatial coordinates of the vortex are assigned to the center of the plaquette in which it was detected.

\newpage
\subsection*{Strain mapping}

The local strain tensor is extracted from the real-space STM topography, $I(\vec{r})$, using a geometric phase analysis (GPA) algorithm \cite{Lawler2010IntraunitcellStates}. This method determines the slowly varying displacement field, $\vec{u}(\vec{r})$, which describes the deviation of lattice points from their ideal reference positions.

The process begins with the selection of two non-collinear Bragg peaks, corresponding to reciprocal lattice vectors $\vec{q}_1$ and $\vec{q}_2$, from the 2D Fourier transform of the STM image. These vectors define the basis of the reference lattice. The real-space image is then demodulated with respect to each of these vectors to create two complex images:
$$
I(\vec{r}) e^{-i \vec{q}_k \cdot \vec{r}}, \quad k \in \{1, 2\}
$$
A Gaussian filter with a standard deviation $\sigma$ is applied to these complex images. This acts as a band-pass filter in Fourier space, isolating the amplitude and phase information associated with the chosen Bragg peaks. The result is a pair of slowly varying complex fields, $T_k(\vec{r})$. The phase of each field, $\phi_k(\vec{r}) = \arg[T_k(\vec{r})]$, is directly related to the component of the local displacement field projected onto the corresponding reciprocal lattice vector:
$$
\phi_k(\vec{r}) = -\vec{q}_k \cdot \vec{u}(\vec{r})
$$
After unwrapping the phase to remove $2\pi$ discontinuities, this pair of equations can be written in matrix form for each real-space position $\vec{r}$:
$$
\begin{pmatrix} \phi_1(\vec{r}) \\ \phi_2(\vec{r}) \end{pmatrix} = - \begin{pmatrix} q_{1x} & q_{1y} \\ q_{2x} & q_{2y} \end{pmatrix} \begin{pmatrix} u_x(\vec{r}) \\ u_y(\vec{r}) \end{pmatrix}
$$
By inverting the $2 \times 2$ matrix of reciprocal lattice vectors, the displacement field components, $u_x(\vec{r})$ and $u_y(\vec{r})$, are solved for at every pixel in the image.

\subsubsection*{Strain tensor calculation}

The calculated displacement field $\vec{u}(\vec{r})$ includes contributions from both lattice strain and experimental artifacts, such as thermal drift and scanner non-linearity. These artifacts typically manifest as a slowly varying background. To isolate the strain-induced component, this background is modeled as a 2D polynomial of a defined order (in this work set to 2) \cite{Wang2025InterplaySuperconductors} and is fit to the displacement field data using a least-squares method. Subtracting this fitted background, $\vec{d}(\vec{r})$, yields the residual displacement field, $\vec{s}(\vec{r}) = \vec{u}(\vec{r}) - \vec{d}(\vec{r})$, which is attributed solely to lattice strain.

The infinitesimal strain tensor, $\varepsilon(\vec{r})$, is then calculated from the spatial derivatives of this residual displacement field:
$$
\varepsilon_{ij}(\vec{r}) = \frac{1}{2} \left( \frac{\partial s_i}{\partial x_j} + \frac{\partial s_j}{\partial x_i} \right)
$$
where $i, j \in \{x, y\}$. The gradients are computed numerically from the discrete displacement field maps.

For better physical interpretation, the strain tensor can be rotated from the Cartesian frame into a basis aligned with a specific crystallographic direction, such as one of the Bragg vectors. The strain tensor is then decomposed into its fundamental components. The trace of the tensor gives the biaxial strain, $\varepsilon_B = \frac{1}{2}(\varepsilon_{xx} + \varepsilon_{yy})$, which represents the fractional change in the local unit cell area. The deviatoric part of the tensor is used to define the uniaxial strain, $\varepsilon_U = \frac{1}{2}(\varepsilon_{xx} - \varepsilon_{yy})$, and the shear strain, $\varepsilon_{xy}$.

Furthermore, the strain tensor is diagonalized at each pixel to find its eigenvalues ($\lambda_1, \lambda_2$) and eigenvectors. The eigenvalues correspond to the principal strains, which are the maximum and minimum extensional strains at that point. The eigenvectors indicate the orthogonal directions along which these principal strains occur. The magnitude of local strain anisotropy is quantified by the absolute difference between the principal strains, $|\lambda_1 - \lambda_2|$. Finally, the magnitude of the gradient of the strain tensor components, $|\nabla\varepsilon_{ij}|$, is calculated to investigate spatial variations in the strain field.

\newpage
\bibliography{references.bib}
\bibliographystyle{unsrt}

\end{document}